\documentclass[longauth]{aa}  
\usepackage{natbib}
\bibpunct{(}{)}{;}{a}{}{,} 
\usepackage[dvipsnames]{xcolor}
\usepackage{graphicx}
\usepackage{xcolor}
\usepackage{amsmath} 
\usepackage{newtxtext,newtxmath}
\newcommand{\rev}[1]{#1 }
\newcommand{\reva}[1]{#1 }
\newcommand{\tmcom}[1]{}

\newcommand{\bbcom}[1]{}

\begin{document}

\title{{Chandra} Survey in the {AKARI} North Ecliptic Pole Deep Field}
\subtitle{Optical/Near-Infrared Identifications of X-ray Sources}
\titlerunning{{Chandra AKARI} NEPD Survey: Source Identifications}
\authorrunning{Miyaji et al.}
\author{
      T.~Miyaji\orcid{0000-0002-7562-485X}\inst{\ref{iae}}
\and  B.~A. Bravo-Navarro\inst{\ref{iae}}
\and  J.~D\'iaz Tello\orcid{0000-0002-0990-0502}\inst{\ref{iae},\ref{prepuc}}
\and  M.~Krumpe\inst{\ref{aip}} 
\and  M. Herrera-Endoqui\orcid{0000-0002-8653-020X}\inst{\ref{iae},\ref{iacu}}
\and  H.~Ikeda\orcid{0000-0002-1207-1979}\inst{\ref{wakayama}}
\and  T.~Takagi\inst{\ref{jsforum}}
\and  N.~Oi\orcid{0000-0002-4686-4985}\inst{\ref{hokkaidojoho}}
\and  A.~Shogaki\inst{\ref{kwangaku}}
\and  S.~Matsuura\orcid{0000-0002-5698-9634}\inst{\ref{kwangaku}}
\and  H.~Kim\inst{\ref{ucla}}
\and  M.~A.~Malkan\orcid{0000-0001-6919-1237}\inst{\ref{ucla}}
\and  H.~S.~Hwang\inst{\ref{snu1},\ref{snu2}} 
\and  T.~Kim\inst{\ref{snu1},\ref{snu2}} 
\and  T.~Ishigaki\inst{\ref{iwate}}
\and  H.~Hanami\inst{\ref{iwate}}
\and  S.~J.~Kim\orcid{0000-0001-9970-8145}\inst{\ref{nthu}}
\and  Y.~Ohyama\orcid{0000-0001-9490-3582}\inst{\ref{asiaa}}
\and  T.~Goto\inst{\ref{nthu}}
\and  H.~Matsuhara\inst{\ref{isas},\ref{sokendai}}
}

\institute{
    Instituto de Astronom\'ia, Universidad Nacional Aut\'onoma
    de M\'exico Campus Ensenada, A.P. 106, Ensenada, BC 22800, Mexico \email{miyaji@astro.unam.mx}\label{iae}
\and
    Instituto de Astronom\'ia, Universidad Nacional Aut\'onoma
    de M\'exico , A.P. 70-264, Ciudad de M\'exico, CDMX 04510, Mexico\label{iacu}
\and
    Preuniversitario UC, Pontificia Universidad Cat\'olica de Chile, Chile\label{prepuc}
\and
  Leibniz-Institut f\"ur Astrophysik Potsdam, An der Sternwarte 16, 14482, Potsdam, Germany\label{aip}
\and
  National Institute of Technology, Wakayama College, 77 Noshima, Nada-cho, Gobo, Wakayama, 644-0023, Japan\label{wakayama}
\and
  Japan Space Forum
   3-2-1, Kandasurugadai, Chiyoda-ku, Tokyo 101-0062 Japan \label{jsforum}
\and
  Space Information Center, Hokkaido Information University,
Nishi-Nopporo 59-2, Ebetsu, Hokkaido 069-8585, Japan \label{hokkaidojoho}
\and
  Department of Physics, Kwansei Gakuin University, 2-1 Gakuen, Sanda, Hyogo 669-1337, Japan\label{kwangaku}
\and
  University of California, Los Angeles, Division of Astronomy \& Astrophysics, 430 Portola Plaza, Los Angeles, CA, 90095-1547, USA\label{ucla}
\and
    Department of Physics and Astronomy, Seoul National University, 1 Gwanak-ro, Gwanak-gu, Seoul 08826, Republic of Korea\label{snu1} 
\and   
    SNU Astronomy Research Center, Seoul National University, 1 Gwanak-ro, Gwanak-gu, Seoul 08826, Republic of Korea \label{snu2}
\and
    Iwate University, 3-18-8 Ueda, Morioka, Iwate, 020-8550, Japan\label{iwate}
\and
    Institute of Astronomy, National Tsing Hua University, 101, Section 2, Kuang-Fu Road, Hsinchu 30013, Taiwan\label{nthu}
\and
    Institute of Astronomy and Astrophysics, Academia Sinica, 11F of Astronomy-Mathematics Building, No.1, Sec. 4, Roosevelt Road, Taipei 10617, Taiwan, R.O.C. \label{asiaa}
\and
    Institute of Space and Astronautical Science, Japan Aerospace Exploration Agency, Sagamihara, 229-8510, Kanagawa, Japan\label{isas} 
\and
    The Graduate University for Advanced Studies, SOKENDAI, Shonan Village, Hayama, Kanagawa, 240-0193, Japan\label{sokendai}
}



\def\LaTeX{L\kern-.36em\raise.3ex\hbox{a}\kern-.15em
    T\kern-.1667em\lower.7ex\hbox{E}\kern-.125emX}

\abstract
{}
{We present a catalog of optical and infrared identifications (ID) of X-ray sources in the {AKARI} North Ecliptic Pole (NEP) Deep field detected with {Chandra} \rev{covering $\sim 0.34\,{\rm deg^{2}}$ with 0.5-2 keV flux limits ranging $\sim 2 \textrm{--} 20\times 10^{-16}\,{\rm erg\,s^{-1}\,cm^{-2}}$.}} 
{The optical/near-infrared counterparts of the X-ray sources are taken from our Hyper Suprime Cam (HSC)/Subaru and Wide-Field InfraRed Camera (WIRCam)/Canada-France-Hawaii Telescope (CFHT) data because these have much more accurate source positions due to their spatial resolution than that of {Chandra} and longer wavelength infrared data. We concentrate our identifications in the HSC $g$ band and WIRCam $K_{\rm s}$ band-based catalogs. To select the best counterpart, we utilize a novel extension of the likelihood-ratio (LR) analysis, where we use the X-ray flux as well as $g - K_{\rm s}$ colors to calculate the likelihood ratio. Spectroscopic and photometric redshifts of the counterparts are summarized. Also, simple X-ray spectroscopy is made on the sources with sufficient source counts.} 
{We present the resulting catalog in an electronic form. The \rev{main ID catalog contains 403 X-ray sources and} includes X-ray fluxes, \rev{luminosities}, $g$ and $K_{\rm s}$ band magnitudes, redshifts and their sources, optical spectroscopic properties, as well as intrinsic absorption column densities and power-law indices from simple X-ray spectroscopy. \rev{The identified X-ray sources \reva{include} 27 Milky-Way objects, 57 type I AGNs, 131 other AGNs, and 15 galaxies. The catalog serves as a basis for further investigations of the properties of the X-ray and near-infrared sources in this field.}}
{We present a catalog of optical ($g$ band) and near-infrared ($K_{\rm s}$ band) identifications of {Chandra} X-ray sources in the {AKARI} NEP Deep field with available optical/near-infrared spectroscopic features and redshifts as well as the results of simple X-ray spectroscopy. During the process, we develop a novel X-ray flux-dependent likelihood-ratio analysis for selecting the most likely counterparts among candidates.}   
  
\keywords{
methods: data analysis -- surveys -- X-ray: galaxies -- galaxies: active  
}

\maketitle

\nolinenumbers
\section{Introduction}

The infrared observatory {AKARI}, which was launched in 2006 and ended its operation in 2011 \citep[e.g.][]{matsuhara05}, has left enormous legacy to the astronomical community, including two-tier dedicated surveys on the North Ecliptic Pole (NEP) region with its InfraRed Camera (IRC) \citep{matsuhara06, wada08, takagi12} consisting of the {AKARI} NEP Deep Field (ANEPD;\,$\sim 0.4\,{\rm deg}^2$) and {AKARI} NEP Wide Field (ANEPW;\,$\sim 5.4\,{\rm deg}^2$) \citep{kimsj12}. The uniqueness of the ANEPD/W field observations lies in the complete coverage of imaging observations with the nine IRC filters, continuously covering 2--25 $\mu$m. The combination of the {AKARI} IRC photometric dataset, combined with other infrared/optical/radio data available in this field, enables us to make Spectral Energy Distribution (SED) decomposition of the sources. In particular, three filters in the 11--18 $\mu$m range filled a gap between {Spitzer}'s Infrared Array Camera (IRAC) and Multiband Imaging Photometer (MIPS) filters, giving unique mid-infrared (MIR) wavelength coverage among deep fields, at least until the Mid-Infrared Instrument (MIRI) makes extensive multi-filter coverage.  Because of the uniqueness of the {AKARI} data, many multi-wavelength follow-up observations have been and are being made. The early Subaru Suprime Cam (SCAM) data concentrated in a sub-region of ANEPD are now superseded by new Subaru Hyper Suprime Cam (HSC) data, covering the entire ANEPW data. The Galaxy Evolution Explorer (GALEX) \citep{burgarella19} and {Herschel} \citep[Photoconductor Array Camera and Spectrometer (PACS)/Spectral and Photometric Imaging Receiver (SPIRE)][]{pearson19,burgarella19} observations have added UV and far infrared coverages respectively.

These data enabled us an IR-selection of AGNs by quantitatively separating the star-formation and AGN torus dust emission component by Spectral Energy Distribution (SED) decomposition \citep[e.g.][]{miyaji19,toba20,wang20}, where the strong mid-infrared (MIR) coverage of {AKARI} was a key. The combination of the strong AGN component in infrared and the weak X-ray (or the upper limit thereof) is a strong tool for identifying highly obscured AGNs. This is because MIR emission is not much affected by the absorption by intervening material, while the X-ray emission, even at $E\ga 2$ keV, is heavily attenuated by Compton-thick (CT) absorbers with column densities of $N_{\rm H}\ga 10^{24}{\rm cm^{-2}}$. Motivated by these, we observed the ANEPD region with our {Chandra} Cycle 12 \citep{miyaji20} proposal. Combined with existing data from the archive, a total of 290 ks of {Chandra} data have been accumulated over $\sim 0.34\,{\rm deg^2}$ and $\sim 450$ X-ray sources have been detected \citep[][hereafter, K15]{krumpe15}.

 While the main purpose of K15 was to present the source detection methods and a source catalog, we also identified Compton-thick AGN candidates from the infrared (IR) AGN luminosity from the SED fits \citep{hanami12} and the X-ray luminosity from {Chandra}.
After the publication of K15, further follow-up observations in the ANEPD and ANEPW fields have been made including those using Subaru HSC (optical) \citep{oi21,kimsj21}, {Herschel} (far infrared) \citep{pearson19,burgarella19}, {GALEX} (ultraviolet) \citep{burgarella19}. Submm and Radio observations have also been made with observatories including the Westerbork Synthesis Radio Telescope (WSRT)\citep{white10} and the Karl G. Jansky Very Large Array (JVLA) (Ishigaki et al. in prep).

Even with the very high spatial resolution ($\la 1^{\prime\prime}$) of {Chandra} near its optical axis, the point spread function (PSF) degrades rapidly with off-axis angle, from sub-arcsecond on-axis to $\sim 5$ arcsec (50\% Energy Encircled Radius) at the off-axis angle of 10 arcminutes
\footnote{\url{https://cxc.harvard.edu/proposer/POG/html/chap4.html}}. Usually for survey observations, including ANEPD, data from the entire field of view (FOV) of the four ACIS-I CCDs ($\sim 16^\prime\times 16^\prime$) are used. The {Chandra} observations of ANEPD are arranged such that the FOVs of 12 (our program) + 2 (archive) observations overlap with neighboring ones. Thus a point in the sky could be observed with multiple off-axis angles with different PSFs. Our source detection algorithms described in K15 take full advantage of the overlapped observations by a simultaneous PSF fitting over overlapped fields. These utilize the {Chandra} version of {\sf emldetect} of the {XMM-Newton} SAS package, simultaneously making multiple PSF fits over the observations with different off-axis angles and energy bands. The positional uncertainty of the source reflects the distribution of the off-axis angles over observations and net counts for each contributing observation. In this method, if there are sufficient photon counts in the observations with small off-axis angles, we get good source positions, even if other observations are at large off-axis angles. However, the sum of photons of all observations contributes to good spectral analysis. But near the edges of the {Chandra} ANEPD, which are only covered at large off-axis angles, the positional uncertainties can be as large as a few arcseconds. Therefore there still are X-ray sources with multiple counterpart candidates.
There are several methods to select the most likely counterpart(s) of a source at a certain wavelength with higher positional uncertainty in catalogs with much lower positional uncertainty. One such method is the likelihood ratio ($LR$) technique developed by \citet{sutherland92} and is frequently used in X-ray surveys \citep[e.g.][]{brusa07,civano12,marchesi16id}. One of the factors of $LR$ is the optical magnitude distribution of X-ray sources $q(m)$. \citet{salvato18} extended the $LR$ to include multi-band photometry and introduced $q(\vec{m})$, which is a function of magnitudes from multi-band photometry. While $q(m)$ or $q(\vec{m})$ can be highly X-ray flux dependent, these authors use a single $q(m)$ or $q(\vec{m})$ for the entire X-ray source population in the respective surveys (or larger surveys at a similar depth). \citet{salvato18,salvato22} used a public code for a general Bayesian-based identification algorithm, NWAY, developed by J. Buchner. While the code allows very flexible prior specifications, X-ray flux dependence of $q(\vec{m})$ has not been included as a part of their prior.   
In this paper, we present a catalog of {Chandra} X-ray source identifications in the {AKARI} NEP Deep Field presented in K15. Due to the positional accuracy, we primarily find X-ray counterparts in the {Gaia}, Subaru HSC-$g$ and Canada-France-Hawaii Telescope (CFHT) Wide-Field InfraRed Camera (WIRCam) $K_{\rm s}$ band catalogs. In selecting the most likely counterpart, we develop the novel X-ray flux-dependent two-band LR technique.
In Sect. \ref{sec:sample}, we explain the source of X-ray, optical-IR imaging, and optical/IR spectroscopy data as well as the sample used in our analysis. In Sect. \ref{sec:id_proc}, we explain our identification procedure, including the X-ray flux-dependent two-band likelihood ratio technique. In Sect. \ref{sec:results}, we present our results including the explanation of our identification catalog, while the catalog itself is published in an electronic version only. Sect. \ref{sec:disc} discusses the results of the identifications. We summarize our work in Sect. \ref{sec:conc}. In Appendix \ref{sec:app_gtc_lbt}, we present our results from our series of multi-object spectroscopic observations using the Gran Telescopio Canarias (GTC)/Optical System for Imaging and low-Resolution Integrated Spectroscopy (OSIRIS) and the Large Binocular Telescope (LBT)/Multi-Object Double Spectrograph (MODS), which are mainly targeted on the AGNs in ANEPD including, but not limited to, the {Chandra} sources presented in Sect. \ref{sec:sample}.
\rev{For luminosity calculations, we use a $\Lambda$-CDM cosmology with 
$H_{\rm 0}=70h_{\rm 70}\mathrm{km\,s^{-1}Mpc^{-1}}$, $\Omega_{\rm m}=0.3$, and 
$\Omega_{\rm \Lambda}=0.7$.}

\section{Observation and data}
\label{sec:sample}

\subsection{{Chandra} observations}

The details of the {Chandra} observations are explained and the point source catalog of the ANEPD is presented in K15. Here we briefly summarize the basic aspects. 

The field was observed with {Chandra} between  December 2010 and April 2011 (cycle 12). Twelve individual ACIS-I pointings with a total exposure time of 250~ks were awarded.  The OBSIDs are 12925, 12926, 12927, 12928, 12929, 12930, 12931, 12932, 12933, 12934, 12935, 12936, and 13244. The central position of the mosaicked observation is roughly R.A.\ = 17h 55m 24s and decl.\ = +66$^{\circ}$ 33$\arcmin$ 33$\arcsec$. In addition, we include two {Chandra} ACIS-I pointings from the archive (10443, 11999), which have observed the southeast corner of the {AKARI} NEP deep field.  The total area covered by our {Chandra} mosaicked survey is $\sim$0.34 deg$^2$. 
This almost completely covers early deep optical and near-infrared images covering 26.3 arcminutes $\times$ 33.7 arcminutes ($\sim$0.25 deg$^2$) with earlier Subaru/SCAM and KPNO/Flamingos respectively.  The observation has been designed to reach an approximately homogeneous coverage of typically $\sim$30--40 ks. In the region with the additional pointings from the archive  (OBSIDs 10443 and 11999), we reach a depth of $\sim$80 ks. In K15, we have made an astrometric correction for each OBSID by cross-matching X-ray source positions with trivial counterparts in the Subaru/SCAM images. \rev{A total of 457 sources are listed in the main source catalog of K15, with flux limits corresponding to 50\% of the survey area are $3\times 10^{-15}$, $8\times 10^{-16}$, and $4\times 10^{-15}\,
{\rm [erg\,s^{-1}\,cm^{-2}]}$ for the 0.5-7, 0.5-2, and 2-7 keV bands respectively. The flux limits on the deepest area are $1\times 10^{-15}$, $2\times 10^{-16}$, and $1\times 10^{-15}\, {\rm [erg\,s^{-1}\,cm^{-2}]}$ for the respective bands.}

\subsection{{Gaia}, Subaru/HSC and CFHT/WIRCam data}
\label{sec:dat_g_ks}

In this paper, we find counterparts from optical and IR data with a higher resolution and positional accuracy than the X-ray data. For this purpose, {Gaia} Data Release 3 (DR3) \footnote{\url{https://www.cosmos.esa.int/web/gaia/}}, our HSC/Subaru and our
WIRCam/CFHT data are used \citep{oi14,kimsj21,oi21}.

 For the HSC data, we use our internal band-merged catalog, which has been produced as a part of the work by \citet{kimsj21}, while those with {AKARI} counterparts are made public along with the paper \footnote{\url{https://zenodo.org/record/4007668\#.X5aG8XX7SuQ}}, the internal catalog includes HSC sources that are not {AKARI} sources. This catalog is slightly updated from the one by \citet{oi21}. Among those HSC band-merged sources, we selected those that are detected in the $g$-band at more than 4$\sigma$ and within our {Chandra} FOV. We choose $g$-band, because there are the largest number of $>4\sigma$ detections among the five HSC bands. The positions of the sources in the band-merged HSC catalogs are forced to be the same in the five $grizy$ bands and thus it is easy to find photometries in other bands. Our WIRCam/CFHT $K_{\rm S}$ band-based catalog has been produced as a part of the work by \citep{oi14}, which also selected by  4$\sigma$ detections at this band.

\begin{figure}
 \centering
 \resizebox{\hsize}{!}{ 
   \includegraphics{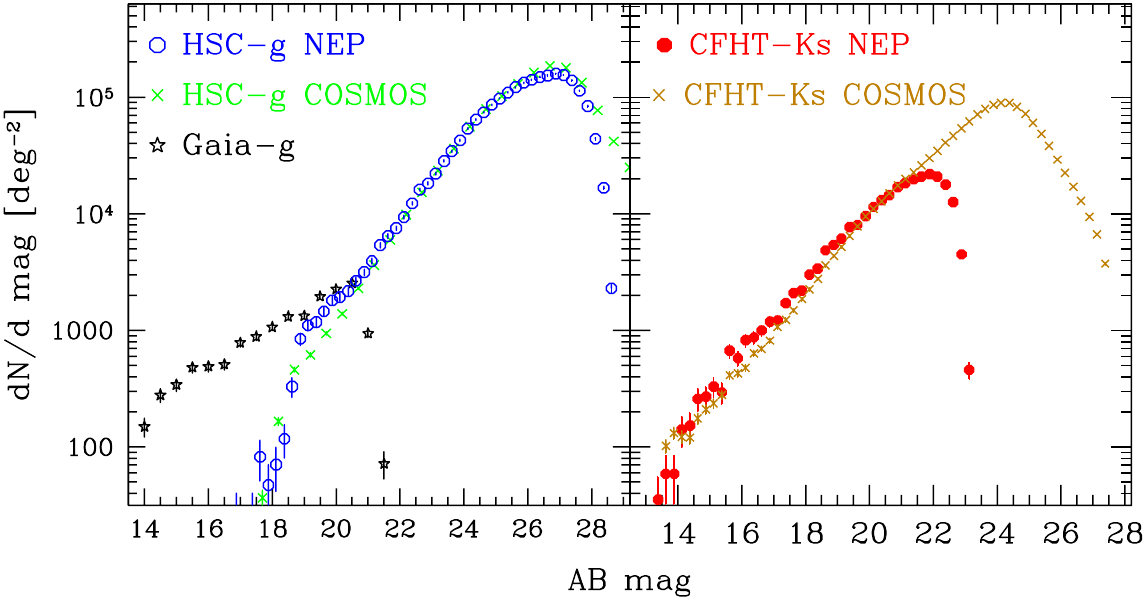}
  }
  \caption{(a) \reva{Number counts} $dN(n)/dm\,{\rm [deg^{-2}\,\,mag^{-1}]}$ of $g$-band sources
    in the {AKARI} NEP Deep Field (within the {Chandra} field, from {Gaia} and {HSC}) and the COSMOS field ({HSC}). (b) \reva{Number counts} for the $K_{\rm s}$ band. Both are from CFHT/WIRCam.    
  }
 \label{fig:nm_comparison}
\end{figure}

 Figure \ref{fig:nm_comparison} shows $g$ and $K_{\rm s}$ band magnitude distributions ($N(m)/dm$) within our {Chandra} FOV. For comparison, the $N(m)$ distributions for the COSMOS field are overplotted for the HSC $g$-band from the HSC-Subaru Strategic Program Data Release 2 (SSP-2) \citep{aihara19}
 \footnote{\url{https://hsc.mtk.nao.ac.jp/ssp/data-release/}} and WIRCam/CFHT $K_{\rm s}$\citep{laigle16}. The $g$ band magnitudes have been corrected for the Galactic extinction.

 The COSMOS data are used as a template for the likelihood ratio analysis described below. We see that the HSC-$g$ band data have similar depths in both fields, while COSMOS has much deeper WIRCam $K_{\rm s}$ data. The HSC $g$-band counts drop suddenly at $g\la 20$ due to image saturations. The {Gaia} catalog, which shows $g$-band magnitudes nicely complements the HSC-$g$ data at the bright end.     

\subsection{Optical and infrared spectroscopy}

Many of the X-ray source candidates have spectroscopic observations in the optical-IR bands from various programs, some are from {AKARI} NEP survey team, others from the archive. The source of spectroscopic data includes Hectspec/Multiple Mirror Telescope (MMT), Hydra/WIYN \citep{shim13},  Deep Imaging Multi-Object Spectrograph (DEIMOS)/Keck from our 2008 and 2011 runs, \citep{churei_mthesis,shogaki_mthesis,kimsj21}, FMOS/Subaru \citep{oi17}, OSIRIS/Gran Telescopio Canarias (GTC, Appendix \ref{sec:app_gtc_lbt}), the SPICY program using the non-slit spectroscopy mode of {AKARI} InfraRed Camera (IRC) \citep{ohyama18}. The results of our newer observations with DEIMOS and  Multi-Object Spectrograph for Infrared Exploration (MOSFIRE/Keck starting 2014 \citep{kimh20}, Hectspec/MMT observed in 2020-2022 \citep{kimt24}, and the large binocular telescope (LBT, Appendix \ref{sec:app_gtc_lbt}) are also included. 

\section{Identification procedure}
\label{sec:id_proc}

\subsection{Selecting counterpart candidates}
\label{sec:selcand}

The X-ray source positions in K15 have been calibrated based on the SCAM data, which are offset
from astrometric solutions of our HSC and CFHT WIRCam data. Before attempting the cross-identification with the HSC and WIRCam sources, we adjust the X-ray source positions by
$\delta_{\rm This Paper}=\delta_{\rm K15}+0.3^{\prime\prime}$, where $\delta$ represents the declination, to match the astrometry of the HSC/WIRCam data. The systematic offset of the R.A. direction between SCAM and HSC/WIRCam data as well as those among {Gaia}, HSC, and WIRCam data are negligible for our purposes ($<0.1^{\prime\prime}$).   

 In K15, we have derived an empirical total positional uncertainty of  
$\sigma_{\rm total} = 5 \times \sqrt{\sigma_{\rm sys.}^2+\sigma_{\rm stat.}^2}$,
where  $\sigma_{\rm stat}$ is the statistical positional uncertainties of our Maximum-likelihood fitting, $\sigma_{\rm sys.}=0.1$ arcsec is the systematic error. Further considering 
the astrometric uncertainties of $\sigma_{\rm astro}=0.2$ arcsec, we define the matching 
radius of $r_{\rm match}=\sqrt{\sigma_{\rm total}^2+\sigma_{\rm astro}^2}$. Note that the notations $\sigma_{\rm total}$ and $\sigma_{\rm astro}$, introduced in K15,  do not represent the standard deviations of any Gaussian. These roughly correspond to 5$\sigma$ in a usual sense. 

We look for optical counterparts in the {Gaia} DR3, HSC/Subaru-$g$ band, and WIRCam/CFHT-$K_{\rm s}$ band selected sources within $r_{\rm match}$. For finding genuine X-ray source counterparts, we use our novel X-ray flux-dependent two-band likelihood ratio
analysis, after excluding special cases. These special cases include those X-ray sources that are part of an extended source and cataloged spuriously as multiple point sources in K15, those sources that are close to the very edge of the combined {Chandra} FOV, and the bright sources that are detected in {Gaia} DR3, a majority of which are saturated in our HSC $g$ band data.
We list the X-ray sources in the category of these special cases in Sect. \ref{sec:results}.

\subsection{X-ray flux-dependent two-band likelihood ratio analysis}

\subsubsection{Formalism}
\label{sec:formalism}

A common procedure for selecting counterparts among multiple candidates is the likelihood ratio (LR) technique, which has been extensively used for finding X-ray source counterparts
\citep[e.g.][]{brusa07,civano12,marchesi16id}, originally developed by \citet{sutherland92} for
the identification of radio sources. The conventional likelihood ratio $LR_{\rm conv}$ is defined by
\begin{eqnarray}
  &LR_{\rm conv}=\frac{q(m)f(r)}{n(m)},\,\nonumber\\
  &\int_{-\infty}^{\infty} q(m)dm=1,\,\,\, \int_0^\infty 2\pi r\,f(r)\,dr=1,
\label{eq:lrconv}
\end{eqnarray}
where $q(m)$ is the apparent magnitude distribution (per mag) of the genuine X-ray source counterparts,
$f(r)$ is the probability density (per solid angle) that the source is at the angle $r$ away from the cataloged X-ray source position, and $n(m)$ is the number density (per solid angle per mag) of the optical sources as a function of apparent magnitude. The likelihood ratio ($LR$) has no units. In practice, $q^\prime=qQ$ and $LR^\prime=LR\,Q$, where $Q$ is the
fraction of the genuine X-ray source counterparts that are within the optical or IR sample. See Sect. \ref{sec:thresh}
for our method to the estimation of $Q$.

Traditionally, for simplicity, $q(m)$ is assumed universal over X-ray sources, independent of their X-ray properties, including the flux. Since the flux of our X-ray sources ranges over four orders of magnitude, we go one step further and consider a flux-dependent likelihood ratio. Furthermore, we analyze two bands of the counterpart candidates simultaneously (HSC-$g$ and WIRCam $K_{\rm s}$ bands). Including these aspects, we   use in this work the following expression for the likelihood ratio:
\begin{equation}
  LR^\prime=\frac{q^\prime(g,K_{\rm s}|S_{\rm X})f(r)}{n(g,K_{\rm s})},\,\,\iint q^\prime(g,K_{\rm S})dg\,dK_{\rm s}=1
\label{eq:lrnew}
\end{equation}
where $q^\prime(g,K_{\rm s}|S_{\rm X})$ is the two-dimensional $g$--$K_{\rm s}$ magnitude distribution of the X-ray source counterparts
given an X-ray flux $S_{\rm X}$. For the normalization of $q^\prime$, the integration limits are defined by the sample.

For those sources with both $g$-band and $K_{\rm s}$ band detections within the magnitude limits of the sample, we assume that $q^\prime(g, K_{\rm s}|S_{\rm X})$ is a product of two Gaussians, one in $A=\log S_{\rm X}+0.4g$ (logarithm of the X-ray to optical flux ratio) and the other in $B=g-K_{\rm s}$, with their mean values of $\bar{A}$ and $\bar{B}$ and the standard deviations $\sigma_{\rm A}$ and $\sigma_{\rm B}$
respectively. The parameters $\bar{A}$, $\bar{B}$, $\sigma_{\rm A}$, and $\sigma_{\rm B}$ can depend on  $S_{\rm X}$.
\begin{eqnarray}
  &q^\prime(g,K_{\rm S}|S_{\rm X})&\propto  \exp\left[-\frac{(A-\bar{A})^2}{2\sigma_{\rm A}^2}\right]\exp\left[-\frac{(B-\bar{B})^2}{2\sigma_{\rm B}^2}\right],\nonumber\\
  & & (A_{\rm min}\leq g \leq A_{\rm max},\,\, B_{\rm min}<B<B_{\rm max})\nonumber  \\
  & &  =  0, \,\,\, (\mathrm{otherwise}).
  \label{eq:q_g_ks}
\end{eqnarray}
The Gaussian components of Eq. \ref{eq:q_g_ks} are truncated at respective $A$ and $B$ values corresponding to the minimum and
maximum $g$ and $g-K_{\rm s}$ magnitudes in the template sample (see below):
\begin{eqnarray}
  A_{\rm min}=\log S_{\rm X}+0.4\,g_{\rm min},\;\; A_{\rm max}=\log S_{\rm X}+0.4\,g_{\rm max}\\
  B_{\rm min}=(g-K_{\rm s})_{\rm min},\;\; B_{\rm max}=(g-K_{\rm s})_{\rm max}
  \label{eq:minmax}
\end{eqnarray}
  
Note the truncation limits of $A$ depend on $S_{\rm X}$. We assume that the center and standard deviation of the Gaussian factor, e.g., for the variable $A$ follow the power-law form of $S_{\rm X}$:
\begin{equation}
  \sigma_{\rm A}=\sigma_{\rm A,14}\,S_{\rm X,14}^{\alpha_{\rm A}}\,,\;\;\;\;\;\;\bar{A}=\bar{A}_{14}\,\,S_{\rm X,14}^{\beta_{\rm A}},
  \label{eq:a_sxdep}
\end{equation}
where $S_{\rm X,14}$ is $S_{\rm X}$ (0.5--7 keV) measured in the units of 
$10^{-14}\,{\rm erg\,s^{-1}\,cm^{2}}$.
We use the forms of $S_{\rm X}$ dependence for $B$ by replacing $A$ in Eq. \ref{eq:a_sxdep} by $B$.  

The motivation of the first Gaussian is the observation that the distribution of X-ray sources in the X-ray flux -- optical flux plane is concentrated along a constant $S_{\rm X}/F_{\rm opt}$ line \citep[e.g., Fig. 11 of][]{marchesi16id} with spread around it.
In reality, the distribution perpendicular to the line is asymmetric and there is a wing toward the lower $S_{\rm X}/F_{\rm opt}$ end, representing host galaxy dominant AGNs and stars. However, if we include proper $g$ magnitude limits in the HSC data, the Gaussian form still gives a good approximation as shown below. As another factor in the distribution, a Gaussian in $B(=g-K_{\rm S})$ is assumed, attempting to be as independent as possible with the $A$ distribution.
A possible correlation between $A$ and $B$ is neglected, which is reasonable for our purpose.

In case the candidate is detected in the $g$-band, but not in the $K_{\rm s}$ band (and vice versa),
we apply the following expressions for $A_g=\log S_{\rm X}+0.4g$ and $A_K=\log S_{\rm X}+0.4K_{\rm s}$:
\begin{eqnarray}
  q^\prime(A_g|S_{\rm X},!K_{\rm s})\propto \exp\left[-\frac{(A_g-\bar{A_g})^2}{2\sigma_{\rm A_g}^2}\right], \nonumber\\
  q^\prime(A_K|S_{\rm X},!g)\propto \exp\left[-\frac{(A_K-\bar{A_K})^2}{2\sigma_{\rm A_K}^2}\right] \label{eq:q_onedet}
\end{eqnarray}
where, $!K_{\rm s}$ and  $\,!g$ mean non-detection in $K_{\rm s}$ and $g$ respectively. We use the forms of $S_{\rm X}$ dependence for $A_g$ ($A_K$) by replacing $A$ in Eq. \ref{eq:a_sxdep} by $A_g$ ($A_K$). The denominator of $LR^\prime$ (c.f. Eq. \ref{eq:lrnew}) should now be $n(g,!K_{\rm s})$, which is the number density per $g$-magnitude of galaxies that are not detected in $K_{\rm s}$, or $n(K_{\rm s},!g)$ defined likewise. 
 For Eqs. \ref{eq:q_g_ks} \& \ref{eq:q_onedet}, the normalization of $q^\prime$ is determined such that the integral over the magnitudes, considering appropriate magnitude limits, becomes unity given $S_{\rm X}$.  

\subsubsection{Estimating $q^\prime$-functions using the COSMOS Legacy template}
\label{sec:qprime}

The $q(m)$ function in Eq. \ref{eq:lrconv} is often estimated by subtracting the magnitude distribution of optical sources around a certain radius ($\sim$ positional errors) around X-ray sources by that of the field scaled by the relative solid angles of the source area and the field.
In our work, however, this method is subject to shot noise at certain regimes, by including X-ray flux dependence and treating two-band optical identifications. Another method is to use an identification catalog of another survey that includes sources with a similar X-ray flux range as our data as a template. Such a survey should have reliable counterparts. For this purpose, we use the source identification catalog of the COSMOS Legacy survey by \citep{marchesi16id}. While they used $LR_{\rm conv}$, the COSMOS Legacy survey has deeper exposures, larger survey area, and highly overlapping ACIS-I exposures \rev{covering a total of $\sim 2.2\,{\rm deg^2}$ within which $\sim 1.7\,{\rm deg^2}$ are uniformly covered at $\sim 160\, {\rm ks}$ of exposure \citep{civano16}. The highly overlapping configuration } makes most of the survey area observed with, on average, smaller off-axis angles. In the COSMOS field, extensive optical-IR dataset, including HSC-$g$ band \citep{tanaka17hsc,aihara19} and $K_{\rm s}$ band from CFHT (and UltraVista) exist, which makes this data convenient for our template. \rev{Typically the offsets between COSMOS Legacy X-ray sources and the counterparts above their $LR$-threshold are $\sim 0.7^{\prime\prime}$}. In constructing our $q^\prime(g, K_{\rm S})$ function, we use the $i$ and $K_{\rm s}$ counterparts of COSMOS Legacy X-ray sources from \citet{marchesi16id}, with the new $g$-band photometry by HSC SSP-2 (Sect. \ref{sec:dat_g_ks}) at the positions of the $i$-band counterpart. We consider an HSC source that is within $0.7^{\prime\prime}$ of the original $i$-band position of the same source. The choice of this separation is that this corresponds to the minimum separation, with very few exceptions, between two different sources in the HSC catalog. We note that, due to the saturation of the HSC data, the brightest sources in $g$-band ($g\leq 19.5$) are not included.

As shown in Fig. \ref{fig:nm_comparison}, the depths of the HSC $g$-band data are similar between the COSMOS Legacy and ANEPD fields, while WIRCam $K_{\rm s}$ data are $\sim 2$ mag deeper in the COSMOS deep field. Thus for the template $q^\prime$-function,
we randomly remove fainter $K_{\rm s}$ detections of the COSMOS Legacy identifications such that the number counts
agree with that of the ANEPD field. Based on the number counts in Fig. \ref{fig:nm_comparison}, we find that
the ratio of the ANEPD and COSMOS number counts can be represented by  $1-0.1025(K_{\rm s}-20.3)^{2.3}$
at $K_{\rm s}>20.3$.  Thus for a COSMOS Legacy ID with $K_{\rm s}>20.3$, a random number $R\,\,(0\leq R \leq 1)$ is generated and
if $R<1-0.1025(K_{\rm s}-20.3)^{2.3}$, we treat this source as undetected in $K_{\rm s}$ for our template purposes. We excluded
those bright sources that are saturated in the HSC image for the template catalog. For the X-ray candidates that are saturated
in the HSC image in our ANEPD field, we use the {Gaia} DR3 catalog and treat them separately.

The parameters of the Eqs. \ref{eq:q_g_ks} and \ref{eq:q_onedet} is determined by a maximum likelihood fitting
of the function over the template ID catalog. The expected number density of objects in the template sample can
be expressed by:
\begin{equation}
  F(g,K_{\rm s},S_{\rm X})=dN(S_{\rm X})/d\log (S_{\rm X})\cdot q(g,K_{\rm s}|S_{\rm X}),
\end{equation}  
where $dN/d\log S_{\rm X}$ is the number of X-ray sources per $\log S_{\rm X}$ as a function of X-ray flux,
and is approximated by a smoothed two power-law form with a shoulder flux  at $S_{\rm X}=S_{\rm X,b}$:
\begin{equation}
  dN/d\log S_{\rm X}\propto \left[(S_{\rm X}/S_{\rm X,b})^{-\gamma_1}+(S_{\rm X}/S_{\rm X,b})^{-\gamma_2}\right]^{-1},
  \label{eq:ns}
\end{equation}
where $dN/d\log S_{\rm X}\propto \gamma_1$ and $dN/d\log S_{\rm X}\propto \gamma_2$ asymptotically at the faint
and bright ends respectively.

\begin{figure}
\centering
\resizebox{\hsize}{!}{\includegraphics{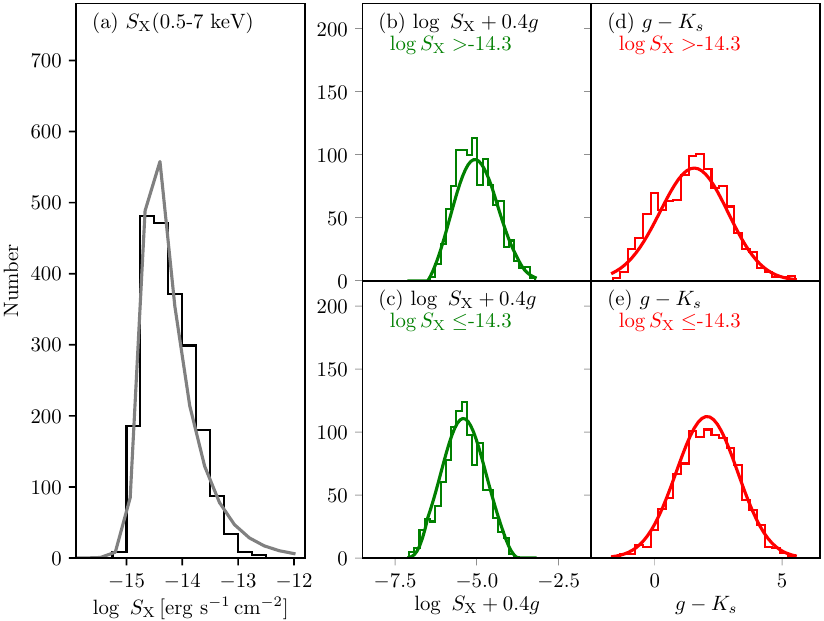}}
   \caption{(a) \reva{Histogram showing} the $dN/d\log S_{\rm X}$ distribution of the template sample of X-ray sources
     that have both $g$ and $K_{\rm s}$, constructed from the COSMOS Legacy survey. The curve shows
     the best-fit model. (b)\&(c) \reva{Histogram} of $\log S_{\rm X}+0.4g$ distribution of the template sample and the best-fit model (curve) for the high and low flux regimes respectively as labeled. (d)\&(e) Same as (b)\&(c) for the $g-K_{\rm s}$ distribution.
     In all panels, $S_{\rm X}$ is the 0.5--7 keV flux in units of ${\rm erg\,s^{-1}\,cm^{-2}}$.}
   \label{fig:comp_2dq}
\end{figure}

In our maximum-likelihood method, \citep[e.g.][]{miyaji15}, we first find best-fit parameter values of the X-ray
number counts Eq. \ref{eq:ns}, i.e., $\alpha_1$, $\alpha_2$ and $\log S_{\rm X,b}$  by minimizing:
\begin{equation}
   {\cal L}_{\rm NS}=-2\sum_i \frac{dN/d\log S_{\rm X,i}}{\int (dN/d\log S_{\rm X})\;d\log\,S_{\rm X}}, 
\end{equation}
where $i$ is over the objects in the template catalog that have both $g$ and $K_{\rm s}$ detections. 
Then the parameters in Eq. \ref{eq:a_sxdep} can be determined by minimizing
\begin{equation}
  {\cal L}_{\rm q}=-2\sum_i \frac{(dN/d\log S_{\rm X,i})q(g,K_{\rm s}|S_{\rm X})}
 {\iiint (dN/d\log S_{\rm X,i})q(g,K_{\rm s}|S_{\rm X}) dK_{\rm s}\,dg\,d\log S_{\rm X}}, 
\end{equation}
in which the parameters of $(dN/d\log S_{\rm X,i})$ are now a fixed.
The best-fit parameters and the minima and maxima of the relevant magnitudes
(Eq. \ref{eq:minmax}) are summarized in Table~\ref{tab:q_params}. The distribution of the objects
in the template catalog is compared with those calculated from the model in Fig. \ref{fig:comp_2dq}
for the X-ray flux, $A=\log S_{\rm X}+0.4g$, and $B=g-K_{\rm s}$. For the latter two, comparisons
are made in low and high X-ray flux regimes as indicated in the figure labels,
where the division is about the median of the template sample.

\begin{figure}
 \centering
 \resizebox{\hsize}{!}{\includegraphics{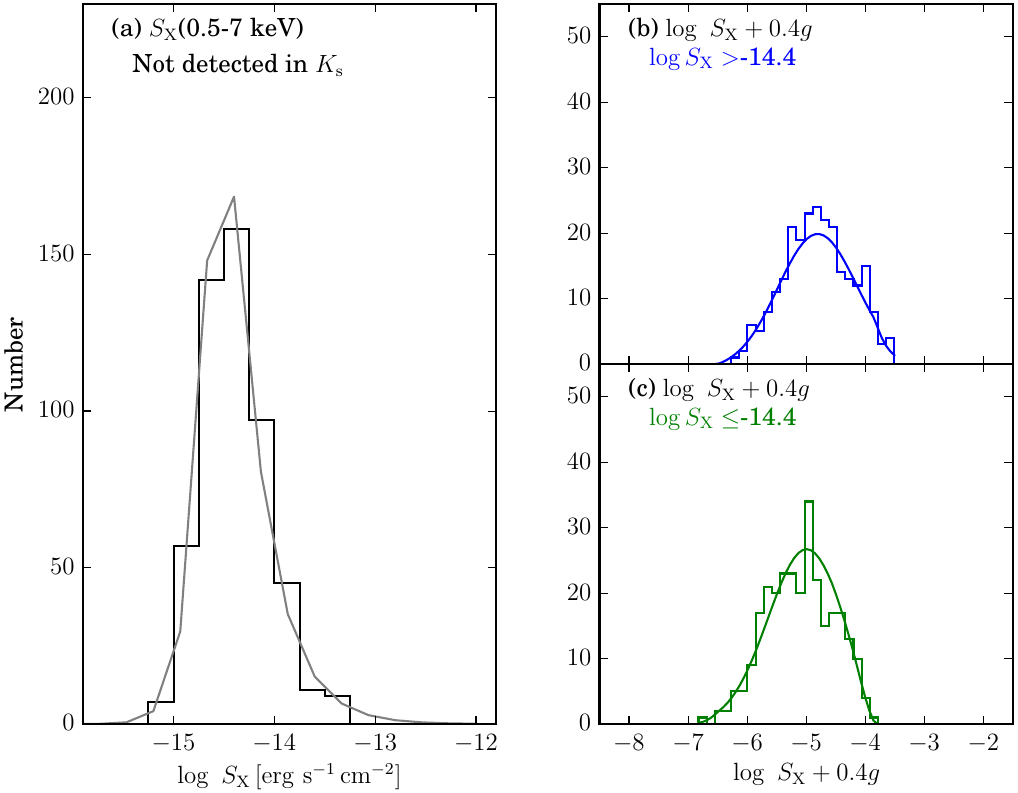}} 
  \caption{(a) \reva{Histogram} showing the $dN/d\log S_{\rm X}$ distribution of the template sample of X-ray sources that are detected in HSC $g$-band, but not in CFHT $K_{\rm s}$ (depth matched with AKARI NEP Deep Field; See text), constructed from the COSMOS Legacy survey. The curve shows the best-fit model. (b)\&(c) \reva{Histogram of} $\log S_{\rm X}+0.4\,g$ distribution
    of the template sample and the best-fit model.
  }
 \label{fig:comp_qg}
\end{figure}

\begin{figure}
 \centering
 \resizebox{\hsize}{!}{\includegraphics{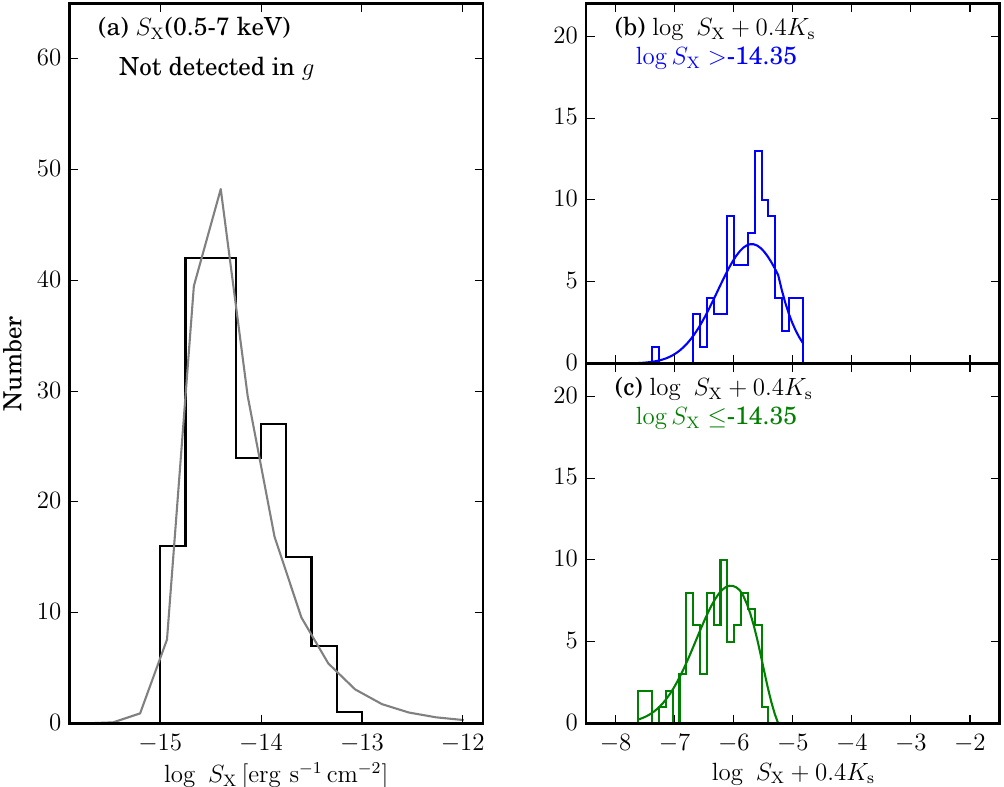}} 
  \caption{Same as Fig. \ref{fig:comp_qg}, except that this is for $K_{\rm S}$ band detected $g$-band
    undetected sources. 
  }
 \label{fig:comp_qk}
\end{figure}

The basic steps are the same as those outlined above for the case of HSC $g$ ($K_{\rm s}$) band-only detection in the template sample. 
Instead of Eq. \ref{eq:q_g_ks}, Eq. \ref{eq:q_onedet} can be used. 
The best-fit parameters for these one-band detection cases are also shown in Table \ref{tab:q_params}. The comparisons of the distributions in the template sample for the one-band detection cases are shown in Figs. \ref{fig:comp_qg} and \ref{fig:comp_qk}.

\begin{table}
  \caption{Best fit Parameters for  $q^\prime(g, K_{\rm s}|S_{\rm X})$, $q^\prime(g|!K_{\rm s},S_{\rm X})$ and $q^\prime(K_{\rm s}|!g,S_{\rm X})$}  
\label{tab:q_params}
\begin{tabular}{lccc}
  \hline\hline
  Parameter & \multicolumn{3}{c}{Values} \\
            & $q^\prime(g, K_{\rm s}|S_{\rm X})$ & $q^\prime(g|!K_{\rm s},S_{\rm X})$ & $q^\prime(K_{\rm s}|!g_{\rm s},S_{\rm X})$ \\ \hline
    $\log S_{\rm X,b}$ & -14.67 & -14.60 & -14.64 \\    
    $\gamma_1$         & 3.90   & 3.25   & 3.52 \\      
    $\gamma_2$         & -0.82   & -1.36  & -0.92 \\
 \\
    $\bar{A}_{14}\,^{a}$ & -5.20  & -4.76  &  -5.70  \\
    $\beta_A\,^{a}$      & -0.03  & -0.03  &  -0.04  \\        
    $\sigma_A\,^{a}$     & 0.70   & 0.65   &   0.55\\
    $\alpha_A\,^{a}$     & 0.016  & 0.005 &   -0.03 \\  
    $\bar{B}_{14}$   & 1.60    & $\cdots$ & $\cdots$ \\
    $\alpha_B$       & -0.18   & $\cdots$ & $\cdots$ \\
    $\sigma_B$       & 1.30    & $\cdots$ & $\cdots$ \\
    $\beta_B$        & 0.060   & $\cdots$ & $\cdots$ \\
 \\
 $g_{\rm min}$,$K_{\rm s, min}$ & 19.51 &  19.75 & 15.79\\
 $g_{\rm max}$,$K_{\rm s, max}$ & 26.38 &  26.43 & 22.75\\
 $(g-K_{\rm s})_{\rm min}$ & -1.67 & $\cdots$ & $\cdots$\\
 $(g-K_{\rm s})_{\rm max}$ &  5.52 & $\cdots$ & $\cdots$\\ \hline
  \end{tabular}
  
$^{a}${The symbol $A$ should be read as $A_g$  for $q^\prime(g, K_{\rm s}|S_{\rm X})$ as well as  $q^\prime(g|!K_{\rm s}S_{\rm X})$
  and $A_K$ for $q^\prime(K_{\rm s}|!g_{\rm s},S_{\rm X})$ respectively.}
\end{table}

\subsubsection{Estimating $LR^\prime$ for the {AKARI} NEP X-ray Source Candidates}

An important factor in $LR^\prime$ is the probability that the X-ray source is at the counterpart's candidate position, and it is normally assumed to be a function of the distance ($r$) between the best-fit X-ray position and that of the counterpart candidate, expressed by $f(r)$.  The source catalog published as a part of K15 has a column {\sf RADEC\_ERR}, which is the statistical 1$\sigma$ error ($\sigma_{\rm stat.}$) of the fitted position projected to an axis. The function $f(r)$ can be expressed by a two-dimensional Gaussian with a standard deviation of $\sigma_{\rm pos} =\sqrt{\sigma_{\rm stat.}^2+\sigma_{\rm sys.}^2}$ with $\sigma_{\rm sys}=0.1^{\prime\prime}$, i.e., $f(r)=\exp\left[-r^2/(2\sigma_{\rm pos}^2)\right]/\left(2\pi \sigma_{\rm pos}^2 \right)$. The astrometric accuracy $\sigma_{\rm astro}=0.2^{\prime\prime}$ is negligible. Figure 8 of K15 shows the comparison of the distribution of the positional deviation between simulated sources and the corresponding detected source positions from the simulated image and that from the assumed $f(r)$, which agree well with each other. For our X-ray source counterpart search, we apply the $LR$ analysis to the HSC-$g$ and WIRCam $K_{\rm s}$ sources within the matching radius $r_{\rm match}$.

The denominator of $LR^\prime$ (Eq. \ref{eq:lrnew}) represents the background density distribution of optical/IR sources as a function of $g$ and $K_{\rm s}$ magnitudes and is proportional to the probability that the X-ray source candidate is a chance coincidence.  We construct $q^\prime(g, K_{\rm s}|S_{\rm X})$, $q^\prime(g|!K_{\rm s}, S_{\rm X})$ and $q^\prime(K_{\rm s}|!g, S_{\rm X})$ from the template sample from the COSMOS Legacy Survey above. On the other hand, we construct $n(g, K_{\rm s})$ from our AKARI NEP dataset. To construct $n(g, K_{\rm s})$ (both $g$ and $K_{\rm s}$ detections), we make a 2-D histogram (in steps of 1 mag. for both bands) of the number density (per unit solid angle) of the combined
HSC-$g$ and WIRCam $K_{\rm s}$ sources per solid angle within the {Chandra} FOV and estimate the value by a 2-D cubic spline fit. For $n(g|! K_{\rm s})$, we spline interpolate a 1-D histogram in $g$ (in steps of 1 mag,) for the HSC $g$-band detected sources without a WIRCam $K_{\rm s}$ band detection. Similarly for $n(K_{\rm s}|!g)$, we spline interpolate a 1-D histogram in $K_{\rm s}$ (in steps of 1 mag,) for the WIRCam $K_{\rm s}$-band detected sources without an HSC $g$ band detection.

\subsection{Determination of Threshold}
\label{sec:thresh}

Once we have the $LR^\prime$ value for each candidate (or $LR=LR^\prime Q$), we can select the most likely counterpart from those candidates
that have $LR>LR_{\rm th}$, where $LR_{\rm th}$ is the threshold.  Some authors take the maximum of $(R+C)/2$ as the threshold \citep[e.g.][]{marchesi16id}, while others \citep[e.g.][]{brusa07,civano12,salvato22} use $R=C$. In this paper, we use $R=C$ to determine $LR_{\rm th}$. 
Before defining $R$ for the whole sample, we define \rev{$R_j$ as the reliability of the $j$-th optical/IR
object that is a candidate counterpart of any X-ray source. This is computed by}
\begin{equation}
  R_j=\frac{(LR)_j}{\sum_i(LR)_i+(1-Q)}=\frac{(LR)_j^\prime Q}{\sum_i(LR)_i^\prime Q+(1-Q)},
  \label{eq:Rj}
\end{equation}
where $Q$ is the fraction of genuine X-ray source counterparts that are in the optical catalog 
introduced in Sect. \ref{sec:formalism}. \rev{The sum is over all the candidates (indexed by $i$)
that are the counterpart candidates of the same of the X-ray source.} 
For the sake of simplicity, we explain the case of  
the conventional single-band optical catalog with a magnitude limit $m_{\rm lim}$ here. In this case, $Q$ can be expressed by
\begin{equation}
  Q=\frac{\int_{-\infty}^{m_{\rm lim}} q(m)\,dm}{\int_{-\infty}^{\infty} q(m) dm}.
  \label{eq:Q_def}
\end{equation}
If the sample has a ``soft'' flux limit, the numerator can be expressed by
$\int q(m) p(m) dm$, where $p(m)$ is the probability that an object with a magnitude $m$ is in the optical sample.

To calculate $Q$ based on Eq. \ref{eq:Q_def}, we need to integrate $q(m)$ to $m=\infty$.
The exact form of $q(m)$ fainter than the magnitude limit is unknown, but it does not have any significance.
However, its integration is important. The $Q$ value can be estimated by the fraction of X-ray sources that
have no candidate ($f_{\rm nocand}$) using the following relation:
\begin{equation}
  N_{\rm X}Q=N_{\rm X}\left(1-f_{\rm nocand}\right)-\sum_k \left(1-\sum_i R_i\right)_k,
  \label{eq:Q_est}
\end{equation}
where $k$ is over the X-ray sources that have at least one candidate, and $i$ is over the candidates for each X-ray source, $j$ is over the candidates of the $k$-th source. The first part on the right-hand side (R. H. S.) is the number of sources that have at least one candidate. The second part (the sum over $k$) represents the number of sources with candidates, where the genuine candidate is not among them.
This can be understood by observing that $R_i$ is the probability of $i$-th candidate to be genuine. 
Note that  $\sum_i R_i$ itself contains $Q$ (Eq. \ref{eq:Rj}). We can estimate $Q$ based on the fraction of X-ray sources that have no candidate and $LR$ of each candidate. We further consider X-ray flux dependence of $Q$ as described in Sect. \ref{sec:thresh_res}.

The reliability over the whole sample is defined by $R=N_{\rm ID}/N_{LR>LR_{\rm th}}$, where $N_{\rm ID}$ is the number of identifications above the threshold, i.e., the sum of all the $R_i$'s of the candidate with $LR>LR_{\rm th}$ over all the X-ray sources and $N_{LR>LR_{\rm th}}$ is the number of all candidates above the threshold. The completeness is defined by $C=N_{\rm ID}/N_{\rm X}$ where $N_{\rm X}$ is the number of X-ray sources, both are functions of $LR$. Once we have $R$ and $C$, we choose our threshold the $LR$ value at $R=C$.

\section{Analysis and results}
\label{sec:results}

\subsection{X-ray sources precluded from the LR analysis}
\label{sec:preclude}

Before attempting the LR analysis, we identify the cataloged X-ray sources (K15 primary catalog) that are not appropriate
for assessing $LR$ (Eq. \ref{eq:lrnew}) values for their candidate counterparts.
As a result, we preclude these sources for the LR analysis.
\begin{itemize}
\item The X-ray sources at the very edge of the combined {Chandra} FOV are excluded. Those are the sources \rev{with a} candidate search area, i.e., a circular region with a radius $r_{\rm match}$ (Sect. \ref{sec:selcand}) from the X-ray nominal position, \rev{that} goes beyond the edge of the FOV. These are ANEPD-CXO392, ANEPD-CXO372 and ANEPD-CXO391.
\item The source detection algorithm of K15 has recognized extended emission from clusters of galaxies as multiple point-like
  X-ray sources. Extended emission from two clusters are recognized in our {Chandra} data \citep{huang21}, one of which is
  previously known as RXJ1757.3+6631 and the other is near our {Chandra} FOV
  at $(\alpha,\delta)=$(17:55:12,+66:33:51).
  The former is catalogued as ANEPD-CXO382, ANEPD-CXO436, ANEPD-CXO370, and ANEPD-CXO311, while the latter as ANEPD-CXO297, ANEPD-CXO257, ANEPD-CXO456, ANEPD-CXO417, ANEPD-CXO208, and ANEPD-CXO302. We do not consider these as true point-like sources but as parts of extended X-ray emission and therefore excluded.
\item \citet{diaztello17} found that ANEPD-CXO104 is an Ultra-Luminous X-ray source (ULX) in one of the spiral arms of a nearby galaxy ($z=$0.027). ANEPD-CXO434 is located at the center of this galaxy. These are also excluded from the $LR$ analysis.
\item Bright sources with $g\leq 19.5$ are mostly saturated in the HSC-$g$ data and they are mostly absent in the HSC catalog. There are a few sources with $g_{\rm HSC}<19.5$ (HSC). However, they are also saturated in the image near the center, and their magnitudes are not reliable.  All of them are, in the {Gaia} DR3 catalog. Due to the very low probability of chance co-incidence, we consider the {Gaia} sources without HSC-$g$ counterpart or $g_{\rm HSC}<19.5$ as secure candidates without calculating $LR$, but assign a very large value ($9.99\times 10^{29}$) in the main identification table.  
\end{itemize}

\subsection{$LR_{\rm th}$ for the sample}
\label{sec:thresh_res}

\begin{figure}
 \centering
 \resizebox{\hsize}{!}{
     \includegraphics{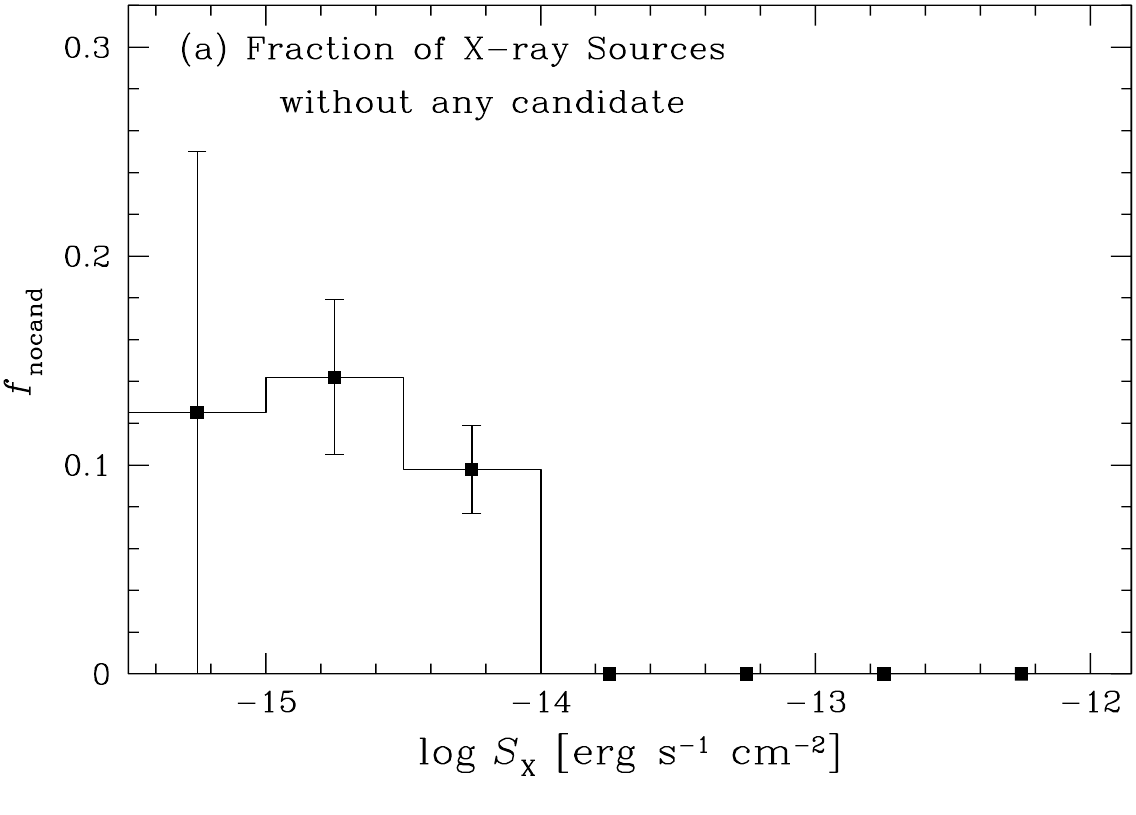}
 }
 
 \resizebox{\hsize}{!}{
     \includegraphics{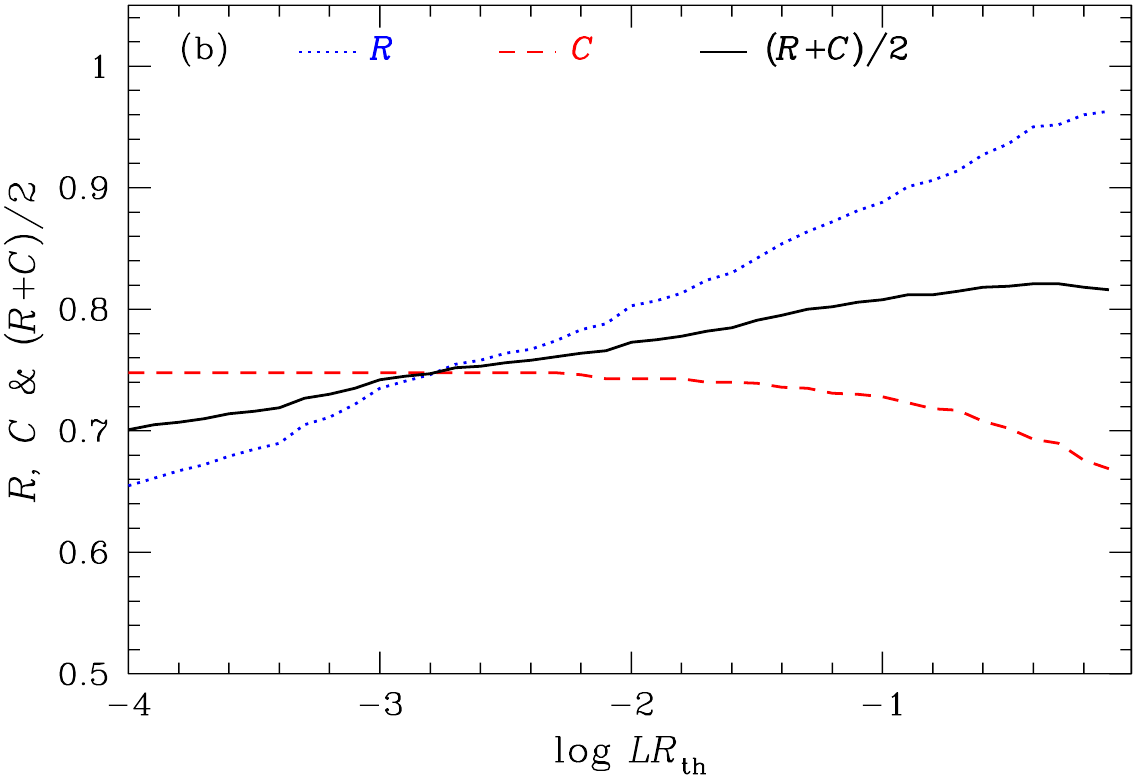}
 }
\caption{(a) Fraction of X-ray sources that have no candidate in any of our 
      HSC-$g$, {Gaia}-$g$ or WIRCam-$K_{\rm s}$
    catalog as a function of $S_{\rm X}$ with the histogram binsize of dex in flux. Approximate 1$\sigma$ errors are also
    shown. (b) Reliability ($R$), completeness ($C$) and $(R+C)/2$ curves as a function of  $\log LR_{\rm th}$ as labeled.}
    \label{fig:nocand_rc}
\end{figure}

We determine the $LR$ threshold following the procedure described in Sect. \ref{sec:thresh}. As a preparation,
we determine the X-ray flux dependent $f_{\rm nocand}$ value to estimate $Q(S_{\rm X})$. Figure \ref{fig:nocand_rc} (a) shows the fraction of our {ANEPD}  X-ray sources that have no candidate as a function of X-ray flux ($N_{\rm nocand}/N_{\rm X}$) where approximate 1$\sigma$ errors roughly estimated by 
$\sqrt{N_{\rm nocand}}/N_{\rm X}$ are also plotted.
We see that there is no such X-ray source at $S_{\rm X}\geq 1\times 10^{-14}\,{\rm [erg\,cm^{-2}\,s^{-1}]}$, while $f_{\rm nocand}$ are consistent with
the average value of 0.11 for all the fainter bins. Using Eq. \ref{eq:Q_est} and \ref{eq:Rj}, where $N_{\rm X}=299$ is the number of
X-ray sources at $S_{\rm X}< 1\times 10^{-14}\,{\rm [erg\,cm^{-2}\,s^{-1}]}$ in our sample,
we find $Q=0.67$. Thus we use $Q=1$ for $S_{\rm X}\geq 1\times 10^{-14}\,{\rm [erg\,cm^{-2}\,s^{-1}]}$ and  $Q=0.67$ for $S_{\rm X}< 1\, \times 10^{-14}\,{\rm [erg\,cm^{-2}\,s^{-1}]}$ respectively.

Figure \ref{fig:nocand_rc} (b) shows the reliability ($R$), completeness ($C$) and $(R+C)/2$ (Sect. \ref{sec:thresh})
as a function of the likelihood ratio threshold ($LR_{\rm th}$). As seen in this figure, $R$ increases, and $C$ decreases with $L_{\rm th}$. Our $(R+C)$ curve is rather flat with no strong peak. Thus in our dataset, it is more beneficial to take the threshold at $R=C$, which is at $\log LR_{\rm th}=-2.8$.   

\subsection{X-ray spectral analysis}
\label{sec:xrayspec}

As a part of the main catalog, we list the results of a simple X-ray spectral analysis for the sources that have net 0.5--7 keV counts of 20 or larger and spectroscopic or photometric redshift measurements. \rev{Out of 403 first candidates listed in Table \ref{tab:main}, 204 have measured extragalactic redshifts, out of which 133 are spectroscopic. Of these 204 sources, we have performed the X-ray spectroscopic analysis of 132 sources.}
The source and background spectra extraction, the creation of the response matrix and ancillary response files, and the combination of the spectra from multiple observations are made in the same way as \citet{miyaji19} for ANEPD-CXO245. The spectral extraction and analysis have been made using the standard tools \reva{from the software packages  {\sf Ciao} (4.11)\footnote{\url{https://cxc.cfa.harvard.edu/ciao/}} and {\sf HEASOFT} (6.25) including {\sf XSPEC} (12.0.1) \footnote{\url{https://heasarc.gsfc.nasa.gov/docs/software/heasoft/}}. The versions of the software packages were the latest as of the first batch of the X-ray spectral analysis.} The model we use for this paper is a simple absorbed power-law form for the photon spectrum: 
\begin{equation}
P(E)\propto {\sf phabs}(N_{\rm H,Gal})*{\sf zphabs}(N_{\rm H,int},z)*E^{-\Gamma},
\label{eq:abspl}
\end{equation}
where $E$ is the photon energy, $N_{\rm H,Gal}=4\times 10^{20}{\rm cm^{-2}}$ is the Galactic absorption column density towards the NEP region \citep{kalberla05}, $N_{\rm H,int}$ is the intrinsic absorption column density of the X-ray source, $z$ is the redshift. The models ${\sf phabs}$ and ${\sf zphabs}$ are the photoelectric absorption without and with redshift respectively as defined in the {\sf XSPEC} package. We first fit with $\Gamma$ as a free parameter. If either of the upper or lower  90\% confidence error exceeds 0.4, we fix $\Gamma$ to =1.8 to obtain $N_{\rm H,int}$ value.  
The main catalog contains columns showing \rev{the values of $N_{\rm H,int}$, $\Gamma$, and the intrinsic 2-10 keV luminosities of the power-law component}. We inspect the spectra for all the X-ray sources that meet the above criteria. If the spectrum significantly deviates from Eq. \ref{eq:abspl}, we \rev{note in the column {\tt COMMENTS\_X} column and in Sect. \ref{sec:comm_obj}}.      

\subsection{Description of identification catalog}
\label{sec:catdesc}
\refstepcounter{table}\label{tab:main}
\refstepcounter{table}\label{tab:suppl}

The main catalog (Table \ref{tab:main}, available electronically at the Strasbourg Astronomical Data Center (CDS)\footnote{\label{foot:cds}\url{https://cds.unistra.fr/}}) lists all the first (largest $LR$) candidates that have $LR\geq LR_{\rm th}$. We also provide a supplementary catalog (Table \ref{tab:suppl}, available electronically at the CDS$^{\ref{foot:cds}}$) showing second and third candidates that have $LR>LR_{\rm th}$ and  $LR$ is larger than 2\% of that of the first candidate. 
In the following,  we describe the columns in the table. The columns included in the supplementary catalog are indicated by '(\#S$N$)', where $N$ is the column number in the supplementary catalog. 



\begin{description}
 \item[\#1(\#S1): {\tt CXONAME}] --- unit: none --- {Chandra} source identification name in the form ANEPD-CXO$NNN$, where $N$ is a one-digit number.  
 \item[\#2: {\tt RA\_X}] --- unit: deg --- right ascension of the X-ray source position
 \item[\#3: {\tt DEC\_X}] --- unit: deg --- declination of the X-ray source position. The declination is offset from that
 in K15 by 0.3$^{\prime\prime}$. (See Sect.~\ref{sec:selcand}.)
 \item[\#4: {\tt RADEC\_ERR\_X}] --- unit: arcsec --- statistical and systematic error ($\sqrt{\sigma_{\rm stat.}+\sigma_{\rm sys}}$)
   of the X-ray source position.
 \item[\#5: {\tt FLUX\_05\_7}] --- unit: ${\rm erg\,s^{-1}\,cm^{-2}}$ --- full band (0.5--7 keV) source flux. \rev{The value of 0 indicates that the source is undetected in this band at the threshold of $ML=9.5$ (K15).}
 \item[\#6: {\tt FLUX\_ERR\_05\_7}] --- unit: ${\rm erg\,s^{-1}\,cm^{-2}}$  --- full band (0.5--7 keV) source flux error (1$\sigma$). \rev{A negative number in this column indicates that the source is undetected and its absolute value is 90\% upper limit.}
 \item[\#7: {\tt FLUX\_05\_2}] --- unit: ${\rm erg\,s^{-1}\,cm^{-2}}$ --- \rev{soft band (0.5--2 keV)} source flux. \rev{See the {\tt FLUX\_05\_7} description.}
 \item[\#8: {\tt FLUX\_ERR\_05\_2}] --- unit: ${\rm erg\,s^{-1}\,cm^{-2}}$ --- \rev{soft band (0.5--2 keV)} source flux error (1$\sigma$). \rev{See the {\tt FLUX\_ERR\_05\_7} description.}
 \item[\#9: {\tt FLUX\_2\_7}] --- unit: ${\rm erg\,s^{-1}\,cm^{-2}}$ --- \rev{hard band (2--7 keV)} source flux. \rev{See the {\tt FLUX\_05\_7} description.}
 \item[\#10: {\tt FLUX\_ERR\_2\_7}] --- unit: ${\rm erg\,s^{-1}\,cm^{-2}}$ --- \rev{hard band (2--7 keV)} source flux error (1$\sigma$). \rev{See the {\tt FLUX\_ERR\_05\_7} description.}
 \item[\#11(\#S2):  {\tt ID\_POS\_REF}] --- Reference of candidate position: \rev{Ga}: {Gaia} DR3, \rev{HSC}: HSC/Subaru, and \rev{Ks}: the $K_{\rm s}$ band. The position is from {Gaia} DR3 source if available. Otherwise from the HSC catalog. If there is no entry in either {Gaia} or HSC catalogs, the $K_{\rm s}$ source position from WIRCAM/CFHT is used. In any case, if the cataloged source positions from different catalogs are within $<0.7^{\prime\prime}$ (see Sect.~\ref{sec:qprime} for this choice), we consider them the same object.  
 \item[\#12(\#S3):  {\tt RA\_ID}] --- unit: deg --- right ascension of the Optical/NIR source position. The position is from { Gaia} DR3 source if available. Otherwise from the HSC catalog. If there is no entries in either { Gaia} or HSC catalogs, the $K_{\rm s}$ source position is used. In any case, if the cataloged source positions from different catalogs are within $<0.7^{\prime\prime}$, we consider them the same object.
 \item[\#13(\#S4): {\tt DEC\_ID}] --- unit: deg --- right ascension of the Optical/NIR position
 \item[\#14(\#S5): {\tt ID\_SEP}(S)] --- unit: arcsec --- separation between the X-ray and Optical/IR positions
 \item[\#15(\#S6): {\tt G\_MAG\_HSC}] --- unit: AB-mag --- HSC-$g$ magnitude. Sources saturated or non-detection in HSC are indicated by the value -9.99. \rev{Sources that are in the HSC band-merged catalog (see Sect. \ref{sec:dat_g_ks}), but not detected in the $g$-band are indicated by the value 99.99.} 
 \item[\#16(\#S7): {\tt G\_MAG\_HSC\_ERR}] --- unit: mag --- Error of HSC-$g$  (1$\sigma$). Sources saturated or non-detection in HSC-$g$ are indicated by the value -9.99. 
 \item[\#17(\#S8): {\tt KS\_MAG\_WI}] --- unit: AB-mag --- WIRCam-$K_{\rm s}$.  Error of HSC-$g$  (1$\sigma$). Sources saturated or non-detection in WIRCam $K_{\rm s}$ are indicated by the value -9.99.
 \item[\#18(\#S9): {\tt KS\_MAG\_WI\_ERR}] --- unit: mag --- Error of WIRCam-$K_{\rm s}$  (1$\sigma$).
   Sources saturated or non-detection in WIRCam $K_{\rm s}$ are indicated by the value -9.99.
 \item[\#19(\#S10): {\tt G\_MAG\_GAIA}] --- unit: AB-mag --- {Gaia} DR3 $g$\_mean. The value -9.99 means non-detection in {Gaia}.
 \item[\#20(\#S11): {\tt GAIA\_MW}] --- boolean --- True: Significant parallax and/or proper motion ($> 3\sigma$) in {Gaia} DR3 indicating a Galactic object, False: No significant parallax, proper motion or {Gaia} source entry.
 \item[\#21(\#S12): {\tt LR}] --- Likelihood ratio
\item[\#22: {\tt SPEC\_Z}] --- Spectroscopic redshift.
\item[\#23: {\tt SPECZ\_Q}] --- Quality of Spectroscopic redshift.  A: the reference source of the redshift has a flag indicating 'excellent/best', B: 'good', and AB: no distinction between these two categories. Normally  'A' is given to those with multiple (three or more) identified features, 'B' to those with two clear emission and/or absorption features (with or without additional weak features), and 'AB' corresponds to B or above. Note that each reference to the redshift has a slightly different definition of the excellent and good quality redshift flags, the divisions among 'A', 'B', and 'AB' are not strict. For the {\tt SPECZ\_REF}=HEC20 objects, the redshift measurements have been made with a cross-correlation technique rather than feature identifications \citep[see][]{kimt24} and we assign AB for these objects. If the redshifts from different observations detecting different features are consistent with one another, we assign 'A' or 'B' based on the combination of the observations following the above guideline. C: One identified feature, D-F: uncertain, P: redshift based on PAH lines. Due to the modeling uncertainties of the broad PAH features, the redshift is subject to small systematic errors ($\delta z/(1+z)<~1\%$). However, since they are based on multiple PAH features, the redshift error is not catastrophic. X: the source literature or database does not give any quality indicator.
 \item[\#24: {\tt SPECZ\_REF}] --- Reference to the spectroscopic redshift. GTC: GTC from Appendix \ref{sec:app_gtc_lbt}, LBT: LBT from Appendix \ref{sec:app_gtc_lbt}, DEI08/DEI11 run: Keck Deimos observations in 2008/2011, \citep{churei_mthesis,shogaki_mthesis}, FMOS12: Subaru FMOS observations in 2012 \citep{oi17}, DEI15/DEI16/MFR17/DEI21: Keck Deimos or Mosfire observations in
   G03 \citep{gioia03}, and R17 \citep{rumbaugh17}.   
\item[\#25: {\tt CLASS}] --- Candidate classifications.  STAR: confirmed stars in the Milky Way by parallax or proper motion from the { Gaia} data. XAGN1: Spectroscopically confirmed broad-line AGN, XAGN: Other AGN 
     XGAL: Galaxies without any sign of AGN. See Table \ref{tab:class} for details. 
 \item[\#26: {\tt Z\_BEST\_OI21}] Photometric redshift from \citet{oi21}.  
 \item[\#27: {\tt MOD\_BEST\_OI21}] The best-fit model template  number from Table~3 of \citet{oi21}.
 \item[\#28: {\tt Z\_NOM}] --- Nominal redshift. Spectroscopic redshift if {\tt SPECZ\_Q} is A, AB, B, C, P, or X.  Photometric redshift by \citet{oi21} otherwise.
\item[\#29: {\tt COMMENTS\_OPTIR}] --- Short comments of optical/IR spectroscopy. Those sources with further comments in Sect. \ref{sec:comm_obj} are marked by an asterisk '(*)'. Classifications in the Baldwin-Phillips-Terlevich (BPT) diagram \citep{bpt} and mid-infrared (MIR) SED studied by \citet{kimh23}, Kim et al. (in prep) for a sub-sample are also included. See Sect.~\ref{sec:kimhclass}.
\item[\#30: {\tt NH22INT}] -- Intrinsic column density at the source redshift in $10^{22}{\rm cm^{-2}}$.
\item[\#31: {\tt NH22INT\_LO}] -- The lower bound of the 90\% confidence range for {\tt NH22\_INT}. 
\item[\#32: {\tt NH22INT\_HI}] -- The upper bound of the 90\% confidence range for {\tt NH22\_INT}. 
\item[\#33: {\tt GAMMA}] -- Photon index $\Gamma$ of the {Chandra} source.   
\item[\#34: {\tt GAMMA\_LO}] -- The lower bound of the 90\% confidence range of {\tt GAMMA}. \rev{The value of -9.99 indicates that $\Gamma$ is fixed during the fit.}
\item[\#35: {\tt GAMMA\_HI}] -- The lower bound of the 90\% confidence range of {\tt GAMMA}.
\rev{The value of -9.99 indicates that $\Gamma$ is fixed during the fit.}
\item[\#36: {\tt LG\_LX\_RAW}] \rev{-- The base-10 logarithm of raw X-ray (0.5-7 keV) luminosity in $h_{70}^{-2}\mathrm{erg\,s^{-1}}$ calculated by $4\pi D_{\rm L}(z)^2 S_{\rm 0.5-7 keV}(1+z)^{\Gamma-2}$, assuming a photon index of $\Gamma=1.4$ for the K-correction (K15). The luminosity distance $D_{\rm L}(z)$ is evaluated at the redshift $z$ from the column {\tt Z\_NOM}. The 0.5-7 keV flux $S_{\rm 0.5-7 keV}$ is from the column {\tt FLUX\_05\_7}. If the source is not detected in the 0.5-7 keV (full) band, the {\tt FLUX\_05\_2}$+${\tt FLUX\_2\_7} value is used instead. Note that the flux column of an undetected band is given the value of zero.}
\item[\#37: {\tt LG\_LX\_INT}] \rev{-- The base-10 logarithm of intrinsic (de-absorbed) X-ray (2-10 keV) luminosity in $[h_{70}^{-2}{\rm erg\,s^{-1}}]$ for those that have available the X-ray spectral analysis results.}
\item[\#38: {\tt LG\_LX\_INT\_LO}] \rev{-- The lower bound of the 90\% confidence range of {\tt LG\_LX\_INT}}.
\item[\#39: {\tt LG\_LX\_INT\_HI}] \rev{-- The lower bound of the 90\% confidence range of {\tt LG\_LX\_INT}}.
\item[\#40: {\tt COMMENTS\_X}] -- Short comments on X-ray spectral analysis. Those sources with further comments in \ref{sec:comm_obj} are marked by an asterisk ('*').
\item[(\#S13): {\tt RANK}] -- (Supplementary catalog only) The {LR} rank. 
\end{description}
Table \ref{tab:class} summarizes the definition and number of sources in each class (\#25 above) in the main catalog.  

\begin{figure}
  \centering
  \resizebox{\hsize}{!}{ 
     \includegraphics{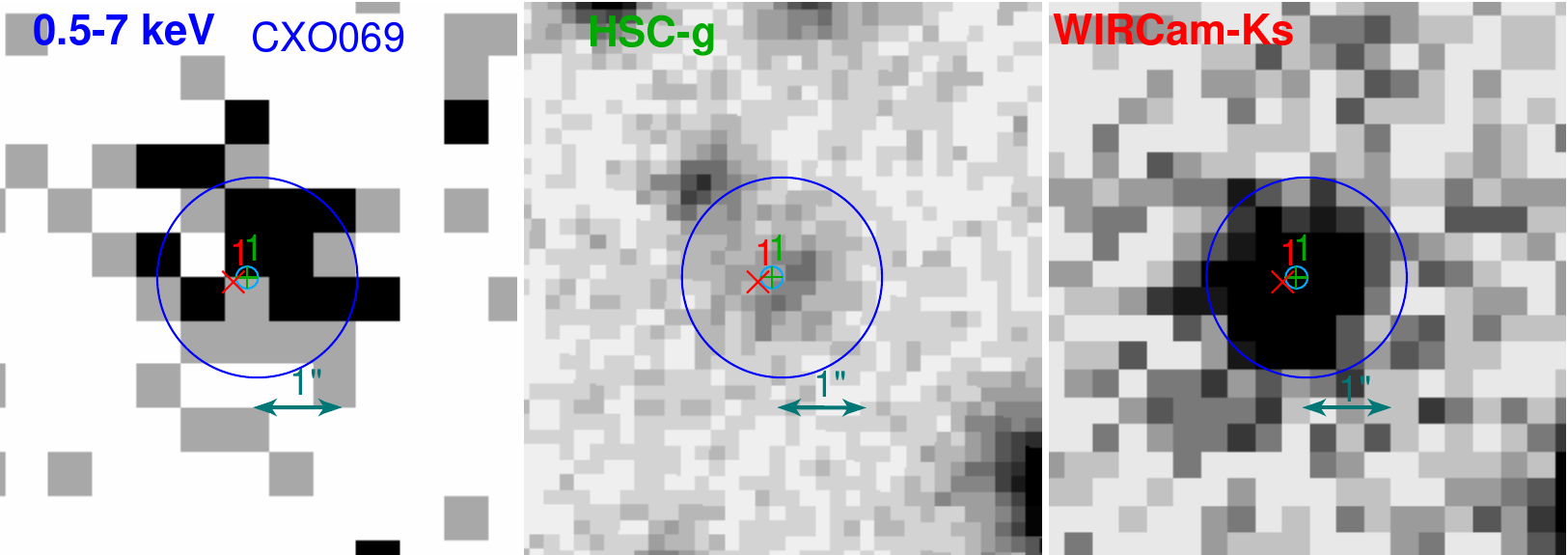}
   }
   \resizebox{\hsize}{!}{ 
     \includegraphics{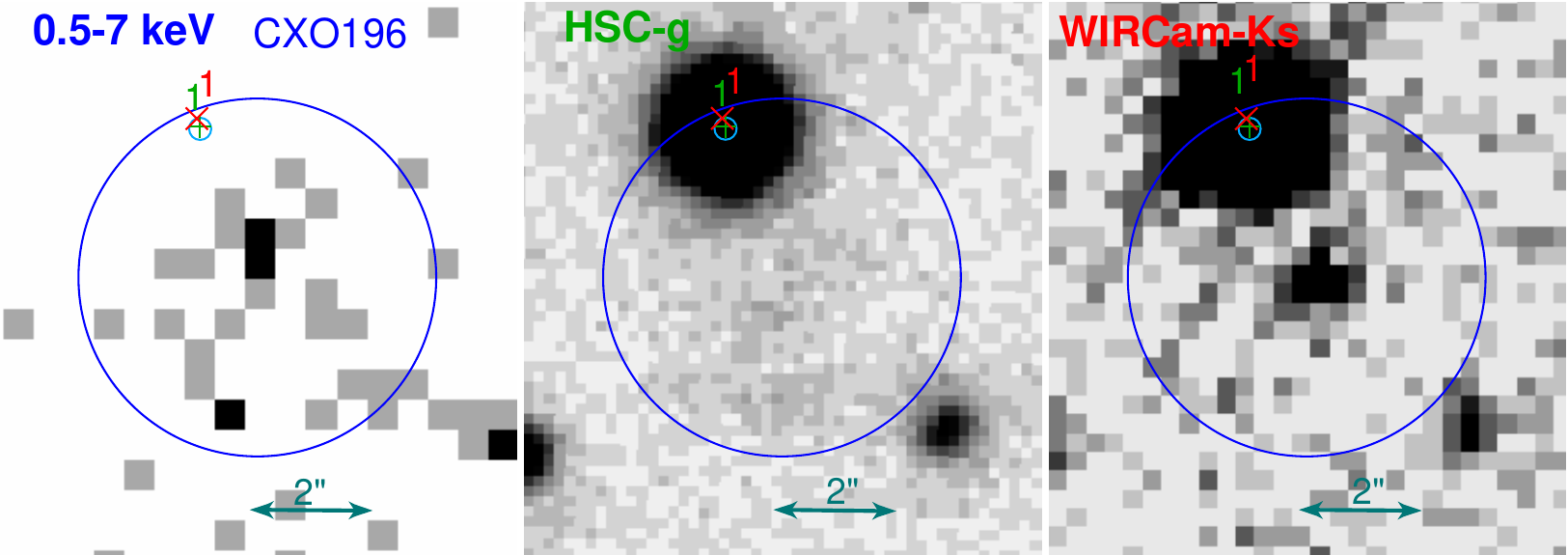}
   }
    \resizebox{\hsize}{!}{ 
     \includegraphics{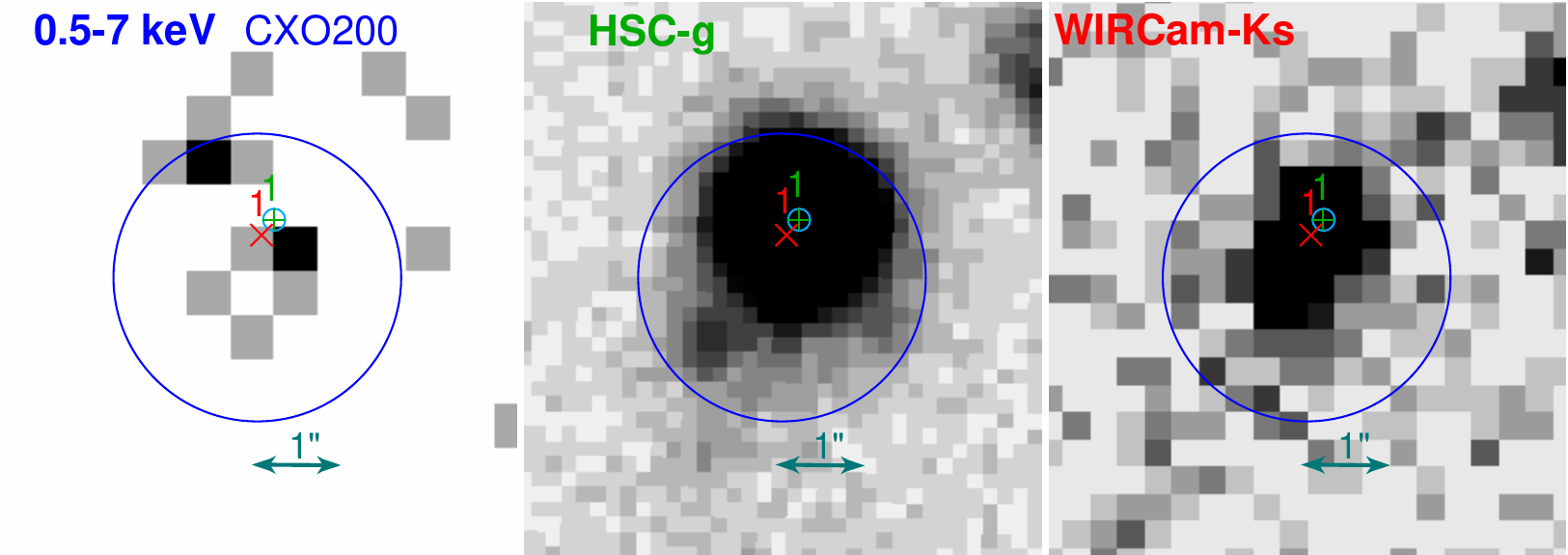}
   } 
 \caption{Examples of poststamp images in X-ray (0.5--7 keV), HSC/Subaru-$g$, and WIRCam/CFHT $K_{\rm s}$ as labeled. In each panel, the blue circle shows the X-ray source matching radius $r_{\rm match}$ (Sect. \ref{sec:selcand}). The positions of the HSC-$g$ and WIRCam $K_{\rm s}$ sources are indicated by a green "+" and red "$\times$" respectively. The cyan circle shows the nominal position of the first ID (largest $LR$). See the description of the table column \#11: {\tt ID\_POS\_REF}. The numbers on the HSC-$g$ and WIRCam $K_{\rm s}$ are the numbers for each source based on the distance from the X-ray source's best-fit position. Full poststamp images are available online\ref{foot:poststamp}.}   
\label{fig:poststamps}
\end{figure}

In addition, the X-ray, HSC $g$-band, WIRCam $K_{\rm s}$ band post-stamp images are available on-line\footnote{\label{foot:poststamp}\url{https://doi.org/10.5281/zenodo.12765949}}. Examples of the post-stamp images are shown in Fig. \ref{fig:poststamps}.

\begin{table}
  \caption{Souce ID Classification}\label{tab:class}
\begin{tabular}{llc}
\hline
 CLASS & Description & Number$^{a}$\\
 \hline\hline
 STAR & \parbox[t]{0.27\textwidth}{Confirmed Milky Way objects that have parallax and/or proper motion
                 detection ($>3\sigma$) from { Gaia} data.} & 27(1)\\ 
\\
 XAGN1 & \parbox[t]{0.27\textwidth}{Spectroscopically confirmed broad-line AGNs that have one or more permitted emission line(s) with $\mathrm{FWHM}>1000\, \mathrm{km\,s^{-1}}$.} & 57(1)\\
 \\
 XAGN  &  \parbox[t]{0.27\textwidth}{Extragalactic sources with spectroscopic redshift measurements without apparent broad line(s) that have either $L_{\rm 0.5-7 keV}>10^{42.5}h_{\rm 70}^{-2} {\rm [erg\,s^{-1}]}$, line ratios in the AGN regime in the BPT diagnostic diagram \citep{bpt} and/or high excitation lines indicative of an AGN.} & 65(2)\\
 \\
 XAGN* & \parbox[t]{0.27\textwidth}{Extragalactic sources that meet the X-ray luminosity criterion for 'XAGN' based on the photometric redshifts.} & 66(2)\\
 \\
 XGAL  &  \parbox[t]{0.27\textwidth}{Extragalactic sources that do not fall into XAGN1/XAGN category based on the spectroscopic redshifts.} & 10(0)\\
 \\
 XGAL* & \parbox[t]{0.27\textwidth}{Extragalactic sources that do not fall into the XAGN* based on the photometric redshifts.} & 5(0)\\
 \\
(...)  &  \parbox[t]{0.27\textwidth}{All others} & 173(14) \\
\hline
\end{tabular}
$^a${\footnotesize The number includes sub-threshold ID's ($LR<LR_{\rm th}$). The number of the sub-threshold ID's is shown in the parenthesis.}
\end{table}

\subsection{Sources with sub-threshold first ID}

Our main catalog contains X-ray source counterparts with $LR<LR_{\rm th}$. These include those that are low-significance X-ray sources, where the best-fit position is probably offset from the true X-ray source position. Other possibilities include the cases where a structure in the nearby background distorts the PSF fit, giving slightly offset best-fit position, and the cases where the true counterpart is not among the $g$-band or $K_{\rm S}$ band catalogs, but an undetected source. Sometimes we see an uncatalogued source in the $K_{\rm S}$ image, with a very faint, also uncataloged source in the HSC-$g$ band close to the X-ray source best-fit position, but another optical source (near the edge of $r_{\rm match}$) becomes the first candidate. An example of such a counterpart is ANEPD-CXO196 (See Fig. \ref{fig:poststamps} and comments in Sect. \ref{sec:comm_obj}.).  

\subsection{Comments on individual sources}
\label{sec:comm_obj}
\begin{description}
    \item[ANEPD-CXO036:] This is a highly absorbed AGN with a borderline Compton-thick absorption. With a simple absorbed power-law model fit (Sect. \ref{sec:xrayspec}), we obtain $\log N_{\rm H}=24.0\,(+0.1;-0.2)\,{\rm cm^{-2}}$ (90\% confidence errors) with $\Gamma=1.8$ (fixed). We make spectral analysis using transmitted, scattered, and torus-reflected \citep[XCLUMPY][]{tanimoto19} components. The results will be reported in a future paper.
    \item[ANEPD-CXO083:] The counterpart has a sub-threshold LR value, but the optical spectrum contains a broad Mg II line indicating AGN. The { CXO} image shows a double peak, indicating that two X-ray sources are blended and recognized as one with a nominal position between two peaks, giving a $\sim 0.9^{\prime\prime}$ offset between the cataloged positions.
    \item[ANEPD-CXO196:] The ID in the catalog is a $g=26.7$ HSC source star ({ Gaia}) near the edge of $r_{\rm match}$ but it is not likely to be the counterpart. There is an uncatalogued source visible in the WIRCam $K_{\rm S}$ band image with no sign of a source in the HSC-$g$ image closer to the X-ray source nominal position. This is likely to be the true counterpart. (See Fig. \ref{fig:poststamps} middle.)
    \item{ANEPD-CXO245:} This is a {bona-fide} Compton-thick AGN with double-peaked narrow lines. Even with only 25 source X-ray counts, we can make reasonable X-ray spectral analysis with transmitted, scattered, and torus-reflected \citep[XCLUMPY][]{tanimoto19} components. See \citet{miyaji19} for detailed discussions about this source. 
    \item[ANEPD-CXO253:] There are two counterpart candidates, both of which have comparably low sub-threshold $LR$, one $7\times 10^{-6}$ at and the other $4\times 10^{-6}$ separated by 1.8$^{\prime\prime}$ and 1.4$^{\prime\prime}$ from the X-ray nominal position respectively. The { CXO} image is weak and has a sign of two peaks. The first candidate has a spectroscopic redshift $z=1.333$ with no sign of AGN. We classify it as CLASS='XAGN' for its X-ray luminosity.
	\item[ANEPD-CXO256:] There are two {Gaia} sources that are identification candidates, one is $g=14.7$, the other $g=17.3$, with separations of 0.5$^{\prime\prime}$ and 0.6$^{\prime\prime}$ from the X-ray source nominal position respectively.
	\item[ANEPD-CXO398:] The counterpart with spectroscopic redshift $z=0.75$ measured by \citet{eckart06} for the same X-ray source (designated by them as CXOSEXSI~J175812.1+664252) is 3$^{\prime\prime}$ away from the X-ray source nominal position and its $LR$ value of $0.33$ is above our threshold. However, our first candidate counterpart with $LR=5.3$, which is fainter ($g$=24.4), is 0.8$^{\prime\prime}$ away from the X-ray source position and has $LR=5.3$. We list our first candidate in Table~\ref{tab:main}.
\end{description}

\section{Discussion}
\label{sec:disc}
\subsection{Spectroscopic and photometric redshifts}

 \begin{figure}
  \resizebox{\hsize}{!}{ 
    \includegraphics{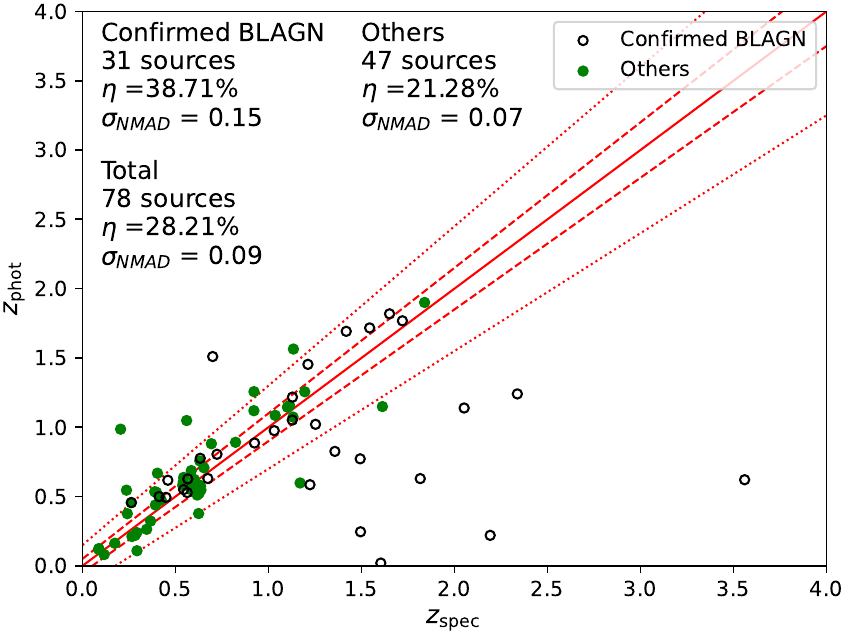}    
  }
 \caption{Comparison between spectroscopic redshift ($z_{\rm spec}$) and photometric redshift ($z_{\rm phot}$) from \citet{oi21} for the 66 sources that have both. Filled circles represent type 1 AGNs (CLASS=XAGN1 in Table 2) while open circles represent non-type 1 objects (CLASS=XAGN/XGAL). The outlier fraction ($\eta$) and NMAD ($\sigma_{\rm NMAD}$) are shown as labels for each of the BL and non-BL objects. The solid red lines represent $z_{\rm spec} = z_{\rm phot}$ and 
 $|z_{\rm phot}-z_{\rm spec} |/z_{\rm spec} = 0.05$ relations, while red dotted lines correspond
to the borders between outlier and non-outlier regimes, corresponding to
$|z_{\rm phot}-z_{\rm spec}|/z_{\rm spec}=0.15$.}
\label{fig:specz_vs_photz} 
 \end{figure}
 
\begin{figure*}
 \centering
  \resizebox{\hsize}{!}{ 
    \includegraphics{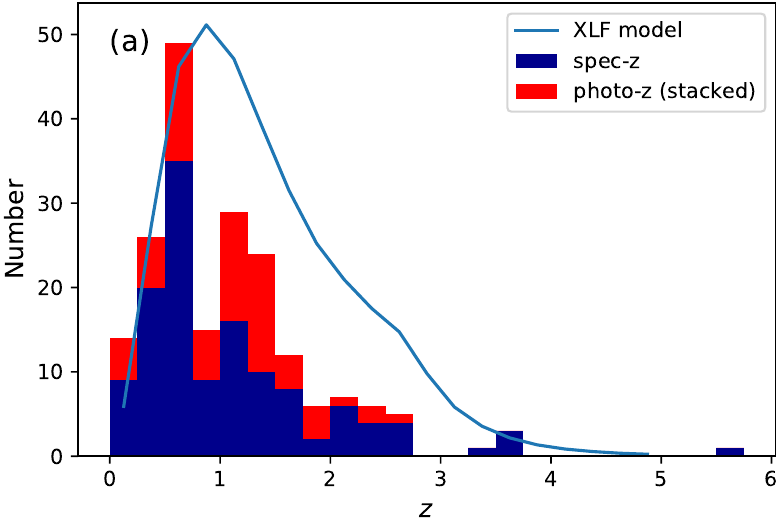}
    \includegraphics{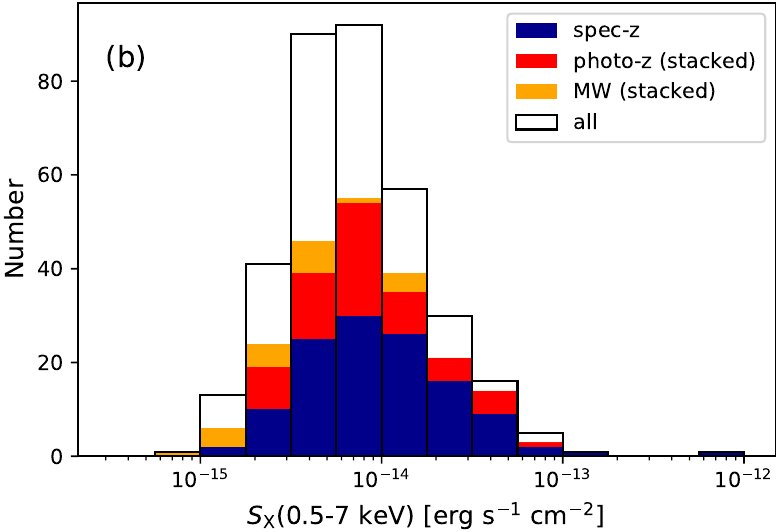}
  }
  \resizebox{\hsize}{!}{ 
    \includegraphics{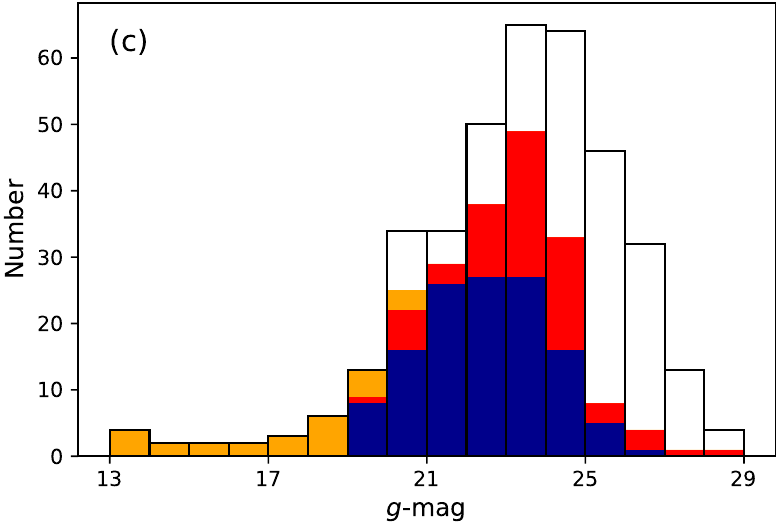}
    \includegraphics{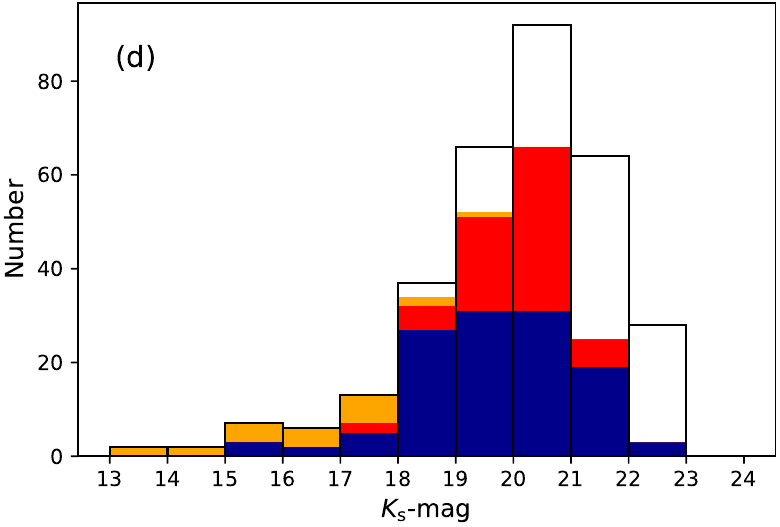}
  }
 
  \caption{(a) Redshift histogram of the spec-z (dark blue histogram) and photo-z (red histogram) samples, where the latter is stacked onto the former. The model expected redshift distribution of AGNs obtained with the X-ray luminosity function model by \citet{miyaji15} for our {Chandra} ANEPD survey is overplotted.  
  (b) \reva{X-ray }(0.5--7 keV) [$S_{\rm X}(0.5-7 {\rm keV})$] histogram of the X-ray sources in our ID catalog. In addition to the histograms for spec-z and photo-z samples, the distribution of Milky Way (MW) objects is also shown (orange histogram, stacked). The distribution of all objects in our sample is also shown as an un-filled histogram.  (c) \& (d) \reva{Histograms of $g$ and $K_{\rm s}$ magnitudeds} for the X-ray sources that are detected in respective bands. The meaning of colors and styles of the histogram are the same as those in panel (b).}
  \label{fig:specz_photz_histo}
\end{figure*}

Among our 376 X-ray source counterparts that have not been classified as Galactic sources, 131  have sufficient quality spectroscopic redshifts ({\tt SPECZ\_Q}=A, AB, B, C, P or X) indicative of extragalactic sources. \citet{oi21} used the L{\small E} P{\small HARE} code \citep{arnouts99,ilbert06} with photometric data in the following 18 optical-IR bands: MegaCam $u^*$, HSC $g$, $r$, $i$, $z$, $y$, WIRCam $Y$, $J$, $K_{\rm s}$, FLAMINGOS $J$, $H$, { AKARI}/IRC $N2$, $N3$, $N4$, {WISE} $W1$, $W2$, and Spitzer/IRAC 3.6 and 4.5 $\mu$ m to obtain photometric redshifts.  In estimating the photometric redshifts, AGN templates are also included. Among our X-ray source counterparts, 78 objects have both spectroscopic and photometric redshifts while 72 objects have photometric redshifts only.

To assess the quality of photometric redshifts for our X-ray-selected AGN sample, we plot photometric versus spectroscopic redshift measurements for the 78 sources that have both. The spectroscopic versus photometric redshift scatter diagram for these 78 objects is shown in Fig. \ref{fig:specz_vs_photz}. Confirmed broad-line (BL) AGNs ({\tt CLASS}= 'AGN1', see Table~\ref{tab:class}) and non-BL AGNs (other extragalactic sources) are marked with different symbols. From these sources, we calculate the normalized median absolute deviation (NMAD), defined by $\sigma_{\rm NMAD}=1.48\times \mathrm{median}[|z_{\rm p}-z_{\rm s}|/(z_{\rm s}+1)]$, where $z_{\rm p}$ and $z_{\rm s}$ are photometric and spectroscopic redshifts respectively. We also define catastrophic failure rate $\eta$, which is conventionally defined as the fraction of objects that have $|z_{\rm p}-z_{\rm s}|/(z_{\rm s}+1)>0.15$.
For all the 78 sources that have both spectroscopic and photometric redshifts, we obtain $\sigma_{\rm NMAD}=0.09$ and $\eta=0.28$. If we exclude the 31 confirmed broad-line AGNs, we find $\sigma_{\rm NMAD}=0.068$ and $\eta=0.213$. For the BL AGNs alone, $\sigma_{\rm NMAD}=0.15$ and $\eta=0.38$.

Due to the possibility of AGN variability, and the featureless continuum of the AGN component, the photometric redshifts of AGNs, especially those of luminous type 1 AGNs with optical-infrared emission dominated by the AGN component, are subject to systematic errors or even catastrophic failures. 
Thus it is natural that these values are worse than those found by \citet{oi21} for the whole optical/infrared selected sample ($\sigma_{\rm NMAD}=0.06$ and $\eta=0.13$ for $z<1.5$), even though AGN templates \citep{salvato09,polletta07} are included.  Figure \ref{fig:specz_vs_photz} shows that a majority of the extreme outliers are $z>1.5$ type 1 AGNs that show lower photometric redshift estimates.  Thus, for our data, the photometric redshift does not have much use for type-I AGN. However, the photometric redshifts give some estimates of the true redshifts of a majority of the non-BL AGNs. Thus the redshift estimates from photometric redshifts of \citet{oi21} are still useful for those that have other indications of obscured (type II) AGNs and highly obscured AGNs (Herrera-Endoqui et al., in prep.). Those include the X-ray sources that have high ratios of the infrared AGN component (obtained from SED fittings) to X-ray (absorption uncorrected) luminosities and those that show absorption in their X-ray spectra.

Figure \ref{fig:specz_photz_histo} (a) shows the redshift distributions of the objects with good spectroscopic redshift measurements with {\tt SPECZ\_Q}=A, AB, B, C, P, or X (spec-z sample) that meet our $LR$ threshold. For those with no good spectroscopic redshift measurement and a photometric redshift measurement is available from \citet{oi21} (photo-z sample), a separate histogram is stacked onto the spec-z sample histogram. To estimate the fraction of X-ray sources that are missing from redshift measurements, we also plot the model-expected redshift distribution of AGNs using the X-ray luminosity function and its evolution from \citet{miyaji15} folded with the real {Chandra} ACIS-I energy response curve and survey solid angle as a function of detection limit in source count rate. The total number of the model-expected AGNs is 352, which is consistent with the number of X-ray sources in our ID catalog presented in Sect. \ref{sec:catdesc} that are not known Milky Way objects. The total number of objects that have good spectroscopic measurements is 126 and among those that do not have good spectroscopic redshift measurements, 83 have photometric redshift estimates. This figure shows that a large fraction of sources at $z>0.6$ do not yet have redshift information.  
We also show histograms of objects that are confirmed Milky Way objects, those that have spectroscopic/photometric redshifts, and those that have no redshift measurements/estimates in various quantities: (b) the X-ray (0.5--7 keV) flux, (c) $g$ band magnitude ({\tt G\_MAG\_GAIA} if available, {\tt G\_MAG\_HSC} otherwise, and (d) $K_{\rm s}$ magnitude.
In panel (b), the expected redshift distribution of the X-ray sources from the X-ray luminosity function by \citet{miyaji15} with the survey solid angle versus 0.5--7 keV band count rate relation of our {Chandra} survey. 

\subsection{Optical and MIR classifications}
\label{sec:kimhclass}

 A comprehensive study for the classifications of galaxies detected by {AKARI} in MIR identified with a Subaru SCAM-based catalog was made by \citet{hanami12} and an early study of the comparison of MIR AGN emission and X-ray finding 28 Compton-thick AGN candidates was presented as a part of K15. We are further conducting revised studies with new datasets including those from HSC, WIRCam, and optical/NIR spectroscopic observations. \citet{kimh23}, Kim et al. (in prep) are conducting a study of calibrating the PAH emission as a star-formation rate indicator in the ANEPD and ANEPW fields. Among the objects in the sample, 40 match within 1 arcsecond of the  {Chandra} X-ray sources ID presented in this paper. Although not all those objects have good-quality optical spectra, we have identified 8 objects that are good enough for detailed spectral classification. In the standard BPT \citep{bpt} diagrams, 7 out of these 8 (88 percent) are classified as ``AGN", with only one being dominated by star formation.  We also identified 40 X-ray-emitting galaxies that have very complete spectral energy distributions. These data include all or nearly all of the 9 {AKARI} mid-infrared bands, which allows us to make a sensitive search for the hot dust continuum known to be produced near a powerful AGN \citep{edelson_malkan12}. Using this more restrictive criterion, we find that 24 of these (60 percent) are identified as having bright AGNs. It is likely that most of the remaining 40 percent also host an AGN, but which is relatively less powerful than its host galaxy, often being classified as a type 2 Seyfert. The BPT classifications and mid-infrared AGNs in this study are indicated in the {\tt COMMENT\_OPTIR} column in Table \ref{tab:main} enclosed by parentheses with 'KM:' followed by the classifications as 'BPTAGN', "BPTSF', 'MIRAGN', and/or 'MIRSF'.  Further studies of the search for infrared AGNs by the SED fits and identifying obscured AGNs by MIR and X-ray comparisons are the subject of a future paper (Herrera-Endoqui et al. in prep).   

\subsection{Current limitations and prospects}

While the {AKARI} IRC photometry and its advantage in finding IR-selected AGNs is a unique feature of the ANEPD field, the supporting multi-wavelength dataset is not as extensive as other premier fields. In particular, the current X-ray data are shallow to take full advantage of the AKARI NEP Deep field data. Thus deeper {Chandra} data are desired. Unfortunately, due to the thermal constraints imposed by the aging of the {Chandra} spacecraft, the efficiency of high ecliptic latitude sky has dropped drastically and this has prevented further {Chandra} observations of this region. 
The NEP region has been selected for one of the deep fields for the recently launched {Euclid}, where full deep slitless spectroscopy in the 0.92-1.85 $\mu$m using both blue and red grisms \rev{\citep[e.g.][]{euclidprepI,euclidprepXX,euclidprepXXXVIII}} will be performed and we eventually can expect redshift measurements of a large majority of the X-ray point sources (as well as {AKARI} infrared selected AGNs) up to $\sim 2.5$. Being a part of the Euclid Wide Survey during the first year of the mission, the depths of optical and near-infrared photometry will reach $I_{\rm E}\sim 26.2$, $Y_{\rm E}$, $J_{\rm E}$, and $H_{\rm E}\sim 24.5$. These will give much improved photometric redshifts.    

\section{Summary}
We present a catalog of optical and near-infrared counterparts of \rev{403 X-ray sources out of 440 X-ray point sources in the {Chandra} data in the {AKARI} NEP Deep Field. These are from the 457 X-ray sources in the main catalog of \citet{krumpe15} after certain exclusions (Sect. \ref{sec:preclude}).} The optical counterparts are searched for among the sources detected in the Subaru HSC $g$-band as well as { Gaia} DR3 catalogs, while near-infrared counterparts are searched for among the sources detected in the CFHT WIRCAM $K_{\rm s}$ band sources. For this purpose, we develop a novel multi-band X-ray flux-dependent extension of the Likelihood Ratio ($LR$) technique. The formulation is developed separately for those that have both $g$ and $K_{\rm s}$ detections and those that have detection in only one of these bands.  The $LR$ threshold for secure identifications is determined by $R=C$, where $R$ is the reliability and $C$ is completeness. 

\rev{Among the 403 first identification candidates, we consider 383 secure in that the $LR$ is above the threshold. The classifications of the 403 identifications are 27 Galactic stars, 132 extragalactic objects with measured spectroscopic redshifts, and 76 extragalactic objects based on photometric redshifts. All the extragalactic sources except 15 objects have AGN luminosities, among which 57 are broad-line (type 1) AGNs. The remaining 173 objects are still unclassified.}  

The identification catalogs are provided in electronic form for the first optical-IR counterpart candidates and the second plus third candidates separately. The catalog also includes the results of the basic X-ray spectral analysis for those that have 20 counts or more in the 0.5--7 keV bands and includes the intrinsic column density and photon index $\Gamma$. 

 We compare the spectroscopic and photometric redshifts for those that have both measurements. For those that have broad emission lines (type I AGNs), the current photometric redshift estimates are not reliable, with $\sigma_{\rm NMAD}\sim 0.16$ and the outlier fraction of  $\eta\sim 39\%$. Excluding those that have broad emission lines, $\sigma_{\rm NMAD}\sim 0.06$ and $\eta \sim 21\%$. The field is a part of the { Euclid} Deep Survey North, where we expect that a very high fraction of the X-ray sources would have spectroscopic redshifts and at least much improved photometric redshifts.      

\label{sec:conc}

\begin{acknowledgements}
 The scientific results reported in this article are mainly based on observations made by the { Chandra} X-ray Observatory (OBSIDs: 12925, 12926, 12927, 12928, 12929, 12930, 12931, 12932, 12933, 12934, 12935, 12936, 13244, 10443, and 11999), {AKARI}, a JAXA project with the participation of ESA and Gran Telescopio Canarias installed at the
Spanish Observatorio del Roque de los Muchachos of the Instituto de Astrof\'isica de Canarias, in the island of La Palma.  This paper uses data taken with the MODS spectrographs built with funding from NSF grant AST-9987045 and the NSF Telescope System Instrumentation Program (TSIP), with additional funds from the Ohio Board of Regents and the Ohio State University Office of Research.
 The authors acknowledge the support by CONACyT Grant Cient\'ifica B\'asica \#252531, 
 UNAM-DGAPA Grants PAPIIT IN111319 and IN114423. TM acknowledges UNAM-DGAPA (PASPA) for support for his sabbatical leave, during which a significant amount of work for this paper has been done. MK is supported by DLR grant FK2 50 OR 2307. HI is supported by JSPS KAKENHI, Grant-in-Aid for Scientific Research (C) 23K03465. HSH acknowledges the support of the National Research Foundation of Korea (NRF) grant funded by the Korean government (MSIT) (No. 2021R1A2C1094577).
 TG acknowledges the support of the National Science and Technology Council of Taiwan through grants 112-2112-M-007 -013, and 112-2123-M-001 -004. We thank Hermann Brunner for his early work in X-ray source detection.
\end{acknowledgements}
\bibliographystyle{aa} 
\bibliography{Miyaji_ChandraAkari_ID_arXiv.bib} 


\begin{appendix}
\section{Gran Telescopio Canarias (GTC) and Large Binocular Telescope (LBT) spectroscopy in the AKARI NEP Deep Field}
\label{sec:app_gtc_lbt}
\subsection{The GTC spectroscopic program}

Given the multiwavelength coverage currently available for the {AKARI} NEP Deep Field, ranging from {Chandra} X-rays to {AKARI} MIR and {Herschel} FIR, we carried out a spectroscopic survey of mid-infrared and X-ray selected AGN. In particular, we aimed to quantify the existence of highly obscured AGNs, including Compton-thick (CT) AGNs. The purpose of the spectroscopy was to obtain accurate redshifts and identify AGN signatures of X-ray and MIR AGNs to help calibrate the $L_{\rm X}/L_{\rm IR,AGN}$ estimate as well as to help the X-ray stacking analysis in the rest frame, to detect the integrated X-ray Fe K-$\alpha$ line, which has large equivalent widths in CT AGNs. Thus, we put priorities on strong CT AGN candidates and type 1 AGN candidates. We also put priorities on targets with high X-ray and/or MIR AGN-component fluxes.

\begin{figure}[h]
\resizebox{\hsize}{!}{ 
    \includegraphics{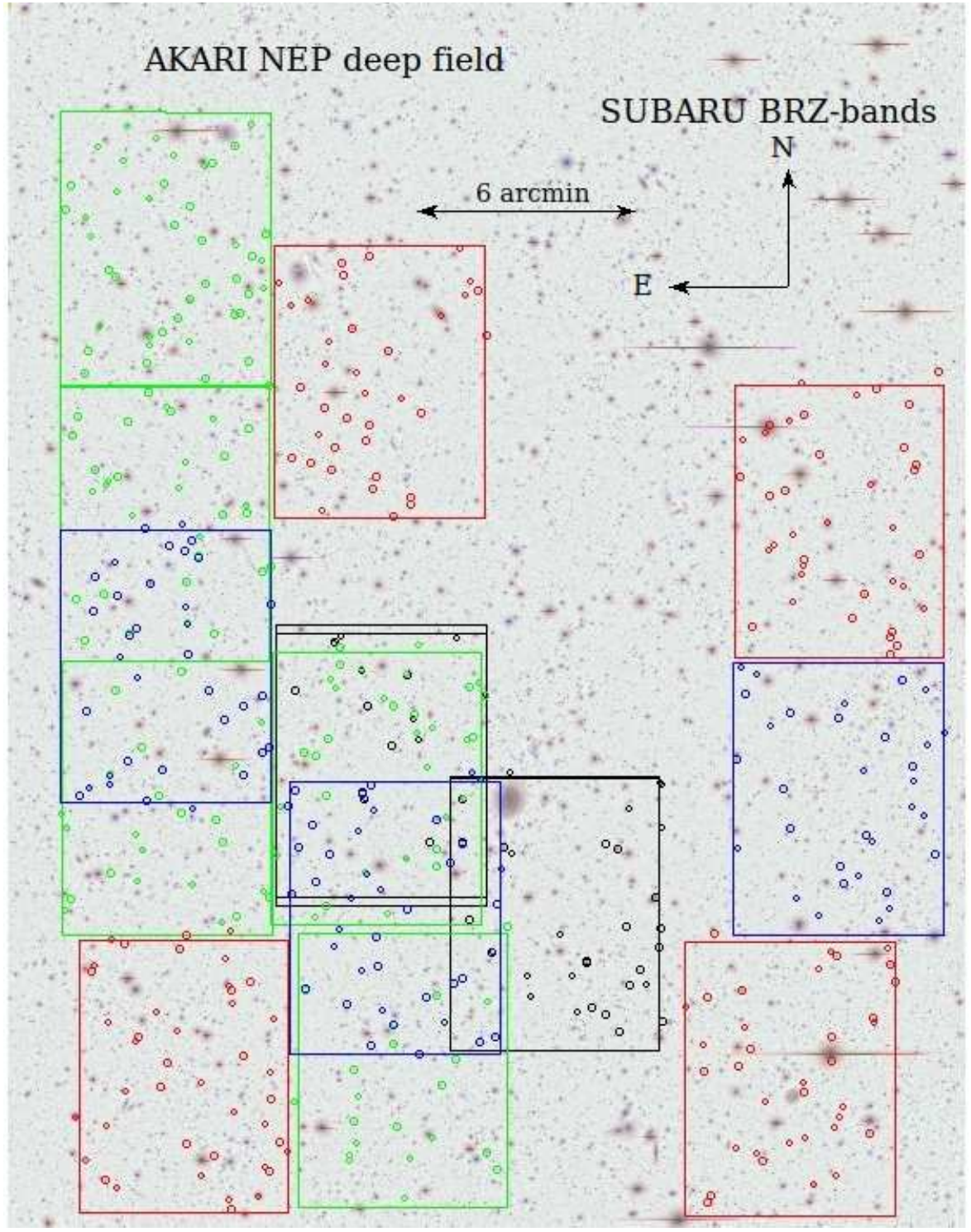}
    }
\caption{Mask positions of OSIRIS/GTC in the {AKARI} NEP Deep Field. Boxes show the OSIRIS coverage area in the different pointings used. Small circles show the objects selected for spectroscopy in each mask. Colors indicate the observation cycle in which each mask and target were used; green for GTC7-14AMEX, red for GTC4-15AMEX, blue for GTC4-15BMEX, and black for GTC4-17AMEX cycles, respectively.}\label{fig:masks}
\end{figure}

Multi-object-spectroscopy of optical counterparts of both {Chandra} X-ray sources as {AKARI} MIR sources were made using the Optical System for Imaging and low-intermediate Resolution Integrated Spectroscopy (OSIRIS) of the Gran Telescopio Canarias (GTC). We obtained the data from one cycle with the long-slit mode performed in 2010 (GTC13-10AMEX) and four observation cycles with the Multi-object spectroscopy mode from 2014 to 2017 (GTC7-14AMEX, GTC4-15AMEX, GTC4-15BMEX and GTC4-17AMEX). 
Considering the Field of View (FOV) of OSIRIS is 7.5 $\times$ 6 arcmin$^{2}$, some of the pointing were located inside the same area already registered to repeat the observation of some objects and also include new ones. This multi-cycle approach was applied to maximize the number of galaxies observed by area and obtain deep observation for the faintest objects but with scientific meaning. Figure \ref{fig:masks} shows the pointings of the masks used in the AKARI NEP Deep Field. In this figure, boxes represent the area covered by OSIRIS, while small circles show the objects observed. Each observation cycle is shown with a different color (GTC7-2014AMEX in green, GTC4-15AMEX in red, GTC4-15BMEX in blue and GTC4-17AMEX in black). The regions close to the center and upper right of the field were not observed due to other spectroscopic campaigns also observing the field, such as DEIMOS/KECK (PI: M. Malkan). We concentrated our observations in the remaining regions of the field but in particular in the area where {Chandra} observations were deeper (bottom left area of the field; see Fig. 1 of K15), pointing several times to that region.

The OSIRIS FOV allows the acquisition of $\sim$30 spectra without overlap between them. In the observations, the R500B and R500R grisms were used for all cycles except the last year (where only R500B grism was utilized), giving dispersions of 3.54 and 4.88 $\AA/{\rm pix}$, respectively. With these grisms, the spectral coverages were from 360 to 700 nm for R500B and from 500 to 1000 nm for R500R, respectively. However, in some cases, the R500B grism practically provided good spectra from 360 to 1000 nm, but with a lower sensitivity toward redder wavelengths, providing thus two measures of features for the same object. The slit width adopted in the observations was 1.2 arcsec, which is a value higher than the average seeing of the whole run (1 arcsec), while the average slit length was 5 arcsec. Typical Integration times of $\sim$700 s and $\sim$1100 s were used for R500B and R500R grisms, respectively, to reach a signal-to-noise ratio (S/N) over 3 in the spectra obtained. Table \ref{tab:obs} summarizes the observation details used in our multi-cycle strategy. Usual calibration frames were acquired during the run plus standard star spectra to make flux calibrations.

The data reduction process was carried out via the \texttt{gtcmos} package\footnote{Developed by Divakara Mayya at INAOE, Puebla, Mexico. Available in \url{https://www.inaoep.mx/~ydm/gtcmos/gtcmos.html}}, which works in IRAF. Before using this package, the OSIRIS images composed of two CCDs with a gap and a slight shift and rotation between them were combined into a single mosaic frame with their pixel coordinates geometrically corrected. This process was performed via the {mosaic2x2$\_$v2} task available on the GTC website. Then, the images were combined and bias subtracted using standard routines of IRAF. The arc line identification, wavelength calibration, and sky subtraction steps were carried out with the {\sf omidentify, omreduce}, and {\sf omskysub} routines, respectively, to obtain the calibrated two-dimensional spectra. The one-dimensional spectra were extracted via the {\sf omextract} routine, while the standard stars were extracted with the {\sf apall} routine. Finally, the flux calibration process was performed with the {\sf standard, sensfunction}, and {\sf calibrate} routines, also within IRAF.

Table \ref{tab:zspec} summarizes the measured spectroscopic redshifts and identified spectroscopic features for the {Chandra} X-ray selected sources and other sources in the ANEPD field observed with OSIRIS/GTC. Each column gives the following information: 
\begin{description}
\item[\#1:] {\tt ObjectID} $-$ Object Identifier adopted from \cite{krumpe15} for {Chandra} X-ray sources. For non-X-ray sources, the ID numbers marked by a '+' is from an internal $z^\prime$-based catalog \citet{hanami12}, '*' from \citet{oi14}, and no mark from \citet{murata13}.
\item[\#2:] {\tt RA} -- unit: deg -- Right Ascension of the object position.
\item[\#3:] \rev{{\tt DEC} -- unit: deg -- Declination of the object position.}
\item[\#4:] {\tt TEL/INSTR} -- Telescope/Instrument
\item[\#5:] {\tt MaskID} -- Mask ID. See Table \ref{tab:obs}.
\item[\#6:] {\tt z\_spec} Spectroscopic redshift estimated for the object.
\item[\#7:] {\tt Q} -- Quality Flag of the measured redshift.
Q indicates the quality of the redshift estimates, discriminating them according to how many features were used to obtain a redshift. The Q flag of "A" means that at least three emission lines were identified (without accounting absorption lines), "B" signifies that only one or two emission lines were recognized (again, without accounting absorption lines), "C" value indicates that only one emission line and no absorption lines were distinguished, and "D" denotes that the redshift estimate is not sufficiently reliable. In addition, the ($^{\star}$) symbol attached to the Q value means that different observations confirm the measured redshift.
\item[\#8:] {\tt CLASS} -- Object Class revealed by the optical spectroscopy.  See Table~\ref{tab:class} for classifications of the X-ray sources. Those that are not detected in X-rays are indicated by the same nomenclature without preceding "X".
\item[\#9:] {\tt SEP} -- unit: arcsec-- Distance between the slitlet position and the catalog object position, from where the ObjectID had its origin.
\item[\#10:] {\tt COMMENTS} -- Comments about the features revealed by the spectrum.\\
\end{description}

\subsection{The LBT spectroscopy}

The LBT observations were obtained with the Multi-Object Double Spectrograph \citep[MODS,][]{pogge10}
instruments. At the time of taking the data, only one of the two main mirrors was equipped with a MODS (specifically the SX version). 
The spectrograph operates in the 3200 to 10,000 {\AA} range with a 6 $\times$ 6 arcminute field of view. Using user-designed laser-machine-cut multi-object masks, the light is split into red and blue-optimized spectrograph channels. The cross-over region is $\sim$5700 {\AA}. For the blue channel, the Grating G400L was used, while the red channel used G670L. The data for both masks, including a total of 37 slits, were collected on May 12 and 13, 2016 with an integration time of 4 $\times$ 1200 s for each mask. 

Standard calibration files for each mask were used. This included biases, flat-field, wavelength-calibration arc lamps, and standard star observations for flux calibration. The data reduction was a two-stage process. First, using \texttt{modsCCDRed}\footnote{\url{https://cgi.astronomy.osu.edu/MODS/Software/modsCCDRed/}}, 
bias and flat-flied corrections were applied. Then, the data were further reduced with scripts written for the reduction of multi-object spectroscopy instruments in \texttt{ESO-MIDAS}. This reduction includes cosmic filtering of the science images, wavelength calibration, weighted extraction of 1D spectra, creation of error spectra, and co-adding the single spectra. 
The log of observations of the LBT spectroscopy is also shown in Table \ref{tab:obs}.
The spectroscopic redshifts and identified features from our LBT observations are also included in Table \ref{tab:zspec}. 

\begin{table*}
\caption{Log of Observations}\label{tab:obs}
\begin{tabular}{lccccc}
\hline
ProgramID & MaskID & R.A. & Decl. & Exp. Time & Date \\
 &   & [deg] & [deg] & [s] & [mmm-dd-yyyy] \\
\hline\hline
GTC13-10AMEX    & long slit & 269.25851 &  66.48038 &  2000 & Jul-08-2010\\
(PI: T. Miyaji) & long slit & 269.21098 &  66.39451 &  2100 & Jul-10-2010\\
                & long slit & 269.31549 &  66.35036 &  2100 & Jul-08-2010\\
                & long slit & 268.61302 &  66.45870 &  2100 & Jul-11-2010\\
                & long slit & 268.38654 &  66.58725 &  2100 & Jul-11-2010\\
                & long slit & 268.65282 &  66.88319 &  2000 & Jun-13-2010\\
                & long slit & 268.67408 &  66.88258 &  2000 & Jun-16-2010\\
\hline                
   GTC7-14AMEX   &   M01B  &  268.94499 &   66.41944 &   3$\times$350  &  May-30-2014\\ 
(PI: T. Miyaji)  &   M01R  &  268.94499 &   66.41944 &   3$\times$510  &  May-30-2014\\ 
                 &   M02B  &  269.21603 &   66.54515 &   3$\times$350  &  Jun-01-2014\\ 
                 &   M02R  &  269.21603 &   66.54515 &   3$\times$510  &  Jun-01-2014\\ 
                 &   M03B  &  268.97550 &   66.54562 &   3$\times$350  &  Jun-01-2014\\ 
                 &   M03R  &  268.97550 &   66.54562 &   3$\times$510  &  Jun-01-2014\\ 
                 &   M04B  &  269.22100 &   66.66619 &   3$\times$350  &  Jun-03-2014\\ 
                 &   M04R  &  269.22100 &   66.66619 &   3$\times$510  &  Jun-03-2014\\ 
                 &   M05B  &  269.22190 &   66.79238 &   3$\times$350  &  Jun-03-2014\\ 
                 &   M05R  &  269.22190 &   66.79238 &   3$\times$510  &  Jun-03-2014\\ 
          \hline   
    GTC4-15AMEX  &   M06B  &  268.97292 &   66.73194 &   3$\times$690  &  Apr-24-2015\\ 
(PI: T. Miyaji)  &   M06R  &  268.97292 &   66.73194 &  3$\times$1130  &  Apr-27-2015\\ 
                 &   M07B  &  269.19500 &   66.41333 &   3$\times$695  &  Apr-29-2015\\ 
                 &   M07R  &  269.19500 &   66.41333 &  3$\times$1130  &  Apr-29-2015\\ 
                 &   M08B  &  268.50208 &   66.41306 &   3$\times$695  &  Apr-26-2015\\ 
                 &   M08R  &  268.50208 &   66.41306 &  3$\times$1130  &  Apr-27-2015\\ 
                 &   M09B  &  268.44208 &   66.66944 &   3$\times$695  &  Apr-28-2015\\ 
                 &   M09R  &  268.44208 &   66.66944 &  3$\times$1130  &  Apr-28-2015\\ 
          \hline   
    GTC4-15BMEX  &   M10B  &  268.44458 &   66.54028 &   3$\times$720  &  Sep-08-2015\\ 
(PI: J. D\'iaz Tello)  &   M10R  &  268.44458 &   66.54028 &  3$\times$1100  &  Sep-08-2015\\ 
                 &   M11B  &  268.95414 &   66.48776 &   3$\times$720  &  Sep-09-2015\\ 
                 &   M11R  &  268.95414 &   66.48776 &  3$\times$1100  &  Sep-09-2015\\ 
                 &   M12B  &  269.21850 &   66.59884 &   3$\times$720  &  Sep-10-2015\\ 
                 &   M12R  &  269.21850 &   66.59884 &  3$\times$1100  &  Sep-10-2015\\ 
          \hline   
    GTC4-17AMEX  &   M13A  &  268.97037 &   66.55417 &   3$\times$915  &  May-24-2017\\ 
(PI: T. Miyaji)  &   M13B  &  268.97037 &   66.55801 &   3$\times$915  &  May-24-2017\\ 
                 &   M14A  &  268.77025 &   66.48832 &   3$\times$915  &  May-27-2017\\ 
                 &   M14B  &  268.77025 &   66.48832 &   3$\times$915  &  May-29-2017\\ 
\hline
LBT             & mask1b & 268.97917 & 66.58167 & 4$\times$1200 & May-12-2016\\
(PI: M. Krumpe) & mask4b & 269.17500 & 66.58167 & 4$\times$1200 & May-13-2016\\
\hline
\end{tabular}
\end{table*}
\begin{table*}    
\caption{Spectroscopic redshifts obtained in our spectroscopic programs with OSIRIS/GTC and LBT/MODS.}\label{tab:zspec}
\begin{tiny}
\begin{tabular}{cccccccccl}
\hline
ObjectID & RA & DEC & OBS/INST & MaskID & $z_{\rm spec}$ &  Q & Class & Dist. & Comments\\
&[deg] & [deg] &        &                &    &      & & [asec] & \\
(1) & (2) & (3) & (4) & (5) & (6) & (7) & (8) & (9) & (10)\\
\hline\hline
ANEPD-CXO105 & 269.30936 & +66.58050 & GTC/OSIRIS & M12B & 1.031 & D & XAGN1 & 0.2 & MgII broad emission line.\\
ANEPD-CXO232 & 269.30613 & +66.54517 & GTC/OSIRIS & M12B & 2.730 & A & XAGN1 & 0.1 & L$\alpha$,SiIV,CIV broad emission lines.\\
ANEPD-CXO146 & 269.30175 & +66.69076 & GTC/OSIRIS & M04B & 0.366 & B & XAGN & 0.0 & K, H absorption lines.\\
ANEPD-CXO077 & 269.29892 & +66.46396 & GTC/OSIRIS & M07B & 0.551 & B & XAGN & 0.3 & [OII] emission, H8, K, H, H$\gamma$ absorption\\
ANEPD-CXO134 & 269.27803 & +66.64869 & GTC/OSIRIS & M12B & 0.617 & D & XAGN & 0.2 & [OII] emission \\
ANEPD-CXO102 & 269.27112 & +66.60539 & GTC/OSIRIS & M12B & 0.306 & D & XAGN & 0.3 & [OII] emission \\
ANEPD-CXO180 & 269.24887 & +66.43140 & GTC/OSIRIS & M07B & 0.627 & B & XAGN & 0.2 & [OII] emission, H8, K, H, H$\delta$ absorption\\
\multicolumn{10}{c}{$\ldots$}\\
\multicolumn{10}{c}{$\ldots$}\\
\multicolumn{10}{c}{$\ldots$}\\
\hline
\end{tabular}
Notes: Seven lines are displayed as examples. The full version of the table is provided electronically at CDS$^{\ref{foot:cds}}$.
\end{tiny}
\end{table*}
\end{appendix}
\end{document}